%% file: ms.tex
\documentclass{pasj01}
\usepackage{color}

\usepackage{dcolumn}

\Received{$\langle$reception date$\rangle$}
\Accepted{$\langle$acception date$\rangle$}
\Published{$\langle$publication date$\rangle$}

\begin{document}

\title{A large sample of shear selected clusters from the Hyper
  Suprime-Cam Subaru Strategic Program S16A wide field mass maps}
\author{Satoshi \textsc{Miyazaki}\altaffilmark{1,2}}
\author{Masamune \textsc{Oguri}\altaffilmark{3,4,5}}
\author{Takashi \textsc{Hamana}\altaffilmark{1,2}}
\author{Masato \textsc{Shirasaki}\altaffilmark{1}}
\author{Michitaro \textsc{Koike}\altaffilmark{1}}
\author{Yutaka \textsc{Komiyama}\altaffilmark{1,2}}
\author{Keiichi \textsc{Umetsu}\altaffilmark{6}}
\author{Yousuke \textsc{Utsumi}\altaffilmark{7}}
\author{Nobuhiro \textsc{Okabe}\altaffilmark{13, 7, 14}}
\author{Surhud \textsc{More}\altaffilmark{5}}
\author{Elinor \textsc{Medezinski}\altaffilmark{8}}
\author{Yen-Ting \textsc{Lin}\altaffilmark{9}}
\author{Hironao \textsc{Miyatake}\altaffilmark{10,5}}
\author{Hitoshi \textsc{Murayama}\altaffilmark{5}}
\author{Naomi \textsc{Ota}\altaffilmark{11}}
\author{Ikuyuki \textsc{Mitsuishi}\altaffilmark{12}}

\altaffiltext{1}{National Astronomical Observatory of Japan, 
 2-21-1 Osawa, Mitaka, Tokyo 181-8588, Japan }
\altaffiltext{2}{SOKENDAI (The Graduate University for Advanced Studies), Mitaka,
  Tokyo, 181-8588, Japan}
\altaffiltext{3}{Research Center for the Early Universe, University of Tokyo,
  Tokyo, 113-0033, Japan}
\altaffiltext{4}{Department of Physics, University of Tokyo, 113-0033, Japan}
\altaffiltext{5}{Kavli Institute for the Physics and Mathematics of the Universe
(Kavli IPMU, WPI), University of Tokyo, Chiba 277-8582, Japan}
\altaffiltext{6}{Institute of Astronomy and Astrophysics, Academia Sinica, Taipei 106 Taiwan}
\altaffiltext{7}{Hiroshima Astrophysical Science Center, Hiroshima University,
  Higashi-Hiroshima Hiroshima 739-8526 Japan}
\altaffiltext{8}{Princeton University Observatory, Princeton  NJ 08544-1001 USA}
\altaffiltext{9}{Institute of Astronomy and Astrophysics, Academia Sinica,
  Taipei 106 Taiwan}
\altaffiltext{10}{Jet Propulsion Laboratory, California Institute of Technology,
  Pasadena CA 91109 USA}
\altaffiltext{11}{Department of Physics, Nara Women's University, Nara 630-8506, Japan}
\altaffiltext{12}{Department of Physics, Nagoya University, Aichi 464-8602, Japan}
\altaffiltext{13}{Department of Physical Science, Hiroshima University, Higashi-Hiroshima, Hiroshima 739-8526, Japan}
\altaffiltext{14}{Core Research for Energetic Universe, Hiroshima University, Higashi-Hiroshima, Hiroshima 739-8526, Japan}

\email{satoshi@naoj.org}


\maketitle

\begin{abstract}
We present the result of searching for clusters of galaxies based on weak
gravitational lensing analysis of the $\sim 160$~deg$^2$ area surveyed by
Hyper Suprime-Cam (HSC) as a Subaru Strategic Program. HSC is 
a new prime focus optical imager with a 1.5 diameter field of view on the
8.2-meter Subaru telescope.  The superb
median seeing on the HSC $i$-band images of $0.56$ arcsec allows the
reconstruction of high angular resolution mass maps via weak lensing,
which is crucial for the weak lensing cluster search. We identify 65
mass map peaks with signal-to-noise (SN) ratio larger than 4.7, and
carefully examine their properties by cross-matching the clusters
with optical and X-ray cluster catalogs. We find that all the 39 peaks
with SN$>5.1$ have counterparts in the optical cluster catalogs, and
only 2 out of the 65 peaks are probably false positives. 
The upper limits of X-ray luminosities from ROSAT All Sky Survey
(RASS) imply the existence of an X-ray under-luminous cluster population. We show
that the X-rays from the shear
selected clusters can be statistically detected by stacking the RASS
images. The inferred average X-ray luminosity is about half that of the
X-ray selected clusters of the same mass.
The radial profile of the dark matter distribution derived from the stacking
analysis is well modeled by the Navarro-Frenk-White profile with a small
concentration parameter value of $c_{500}\sim 2.5$, which suggests
that the selection bias on the orientation or the internal structure for
our shear selected cluster sample is not strong.  
\end{abstract}

\section{Introduction}

Clusters of galaxies are important tools for testing theories of structure
formation and constraining cosmological parameters via the abundance and its
evolution.  Traditionally, clusters have been identified using their
optical or X-ray emission and thus the resulting catalogs are inevitably 
biased toward the selection of more luminous objects. In the dark
matter dominated universe, however, it is challenging to quantify the
selection function of such catalogs, given the uncertainty in the
connection between dark and luminous objects. Ideally we want to
select clusters of galaxies based on their masses, which are the the
most critical parameter in characterizing clusters. A mass selected
sample of clusters allows more direct and straightforward comparisons
with $N$-body simulations, which leads to better understanding of the
connection between luminous and dark matter in clusters as well as
more direct tests of structure formation theories.

\begin{longtable}{*{7}{l}} \caption{Attempts to search for clusters based on the
  weak lensing shear\label{table:prevpeak}} \hline
Telescope & $n_{\rm g}$   & Median Seeing & Area &  $n_{\rm peak}$ ($\nu>4.5$) & Purity & Ref.\\
          & [arcmin$^{-2}$] & [arcsec]  & [deg$^2$] & [deg$^{-2}$] & [\%] & \\ \hline
Mayall    & 19 & 0.90 & 7.5 & 0.27 & 100 (2$/2$) & \citet{2006ApJ...643..128W}\\
CFHT      & 35 & 0.91 & 3.6 & 0.56 & 100 (2$/$2) & \citet{2007A&A...462..459G}\\
MPG$/$ESO & 12 & 0.90 & 19 & 0.89 & 65 (11$/$17)$^{*}$& \citet{2007A&A...462..875S} \\ 
Subaru    & 46 & 0.55 & 2.2& 0.95 & 75 (3$/$4)& \citet{2007ApJ...669..714M} \\
CFHT   &  11.2 & 0.71 & 55.0 & 0.93 & 59 (30$/$51) & \citet{2012ApJ...748...56S}
\\
Subaru & 34.5 & 0.71  &$\sim 3$& 2.3&100(7/7)&Utsumi et al. (2014) \\
Subaru & 24 & 0.57 & 9.0 & 0.89 & 100 (8/8) & \citet{2015PASJ...67...34H} \\
Subaru &  20.9 & 0.58 & 2.3  & 3.5  & 100 (8$/$8)& \citet{2015ApJ...807...22M} \\
\hline
\endhead
\hline
\endfoot
\multicolumn{7}{l}{$^*$only $S$-statistics peaks are considered in this compilation.} \\ %\hline
\endlastfoot
\\ \end{longtable}

Weak gravitational lensing has the potential to offer the most reliable
route to select clusters independently of any assumptions about their
baryon contents and thermal and dynamical states. This technique uses
statistical measures of distorted shapes (``shear'') of faint distance
galaxies induced by the tidal gravitational field integrated along the
line-of-sight. This, in turn, allows a reconstruction of two-dimensional 
surface mass density maps \citep{1993ApJ...404..441K,seitz95,1996MNRAS.283..837S} 
where we are able to locate the dark matter concentrations from the
peak positions. The redshift information can be obtained by correlating
the peak location with optically selected cluster catalogs. Because
the selection function of the peaks can be defined in a straightforward 
way \citep{2012MNRAS.425.2287H}, such shear-selected cluster catalog
is useful for both cluster astrophysics and cosmology. For example,
the shear selected cluster sample will complement studies planned by eRosita
all sky X-ray cluster surveys through the comparison of the cluster selection
functions.

While the weak lensing search of clusters has several advantages as
discussed above, it has not been possible to construct a large sample of
shear selected clusters due to observational difficulties. First, the
demand on the observing facility is high. In order to locate the
cluster scale dark matter halo via weak lensing, deep imaging with
the limiting magnitude deeper than $i_{\rm AB}\;\sim\;24.5$ is
necessary to achieve sufficient number density of source galaxies. 
At the same time, wide field imaging camera is necessary to search for 
rare objects like clusters. Satisfying both the depth and wide area
for constructing a large sample of shear selected clusters has been
difficult in previous imaging surveys. Second, because the lensing
kernel is broad along the line-of-sight direction, the two-dimensional
lensing signal is affected by small foreground and background
structures along the line-of-sight, which may cause ``fake'' signals that
are not associated with single massive objects \citep{white02, henna05}.  
Such contaminations by projections may reduce the value of shear selected
cluster catalogs. For example, \citet{hamana04} evaluated the
contamination rate of shear selected clusters based on the weak
lensing ray-trace simulation and found that the contamination rate can
reach up to 42\% when one sets the threshold of the signal-to-noise
ratio (SN, or $\nu$) to 4. Adopting a higher threshold for SN mitigates
such contaminations, although this further reduces the number of shear selected
clusters.

Observationally, there have been several attempts to search for clusters via
weak lensing (see also Table~\ref{table:prevpeak}). The Deep Lens
Survey is one of the pioneering projects where
\citet{2001ApJ...557L..89W} announced the first discovery of a galaxy
cluster at $z=0.276$ by weak lensing. \citet{2007A&A...462..875S} identified 17
mass map peaks with $\nu>4.5$ on 19~deg$^2$ area imaged by MPG/ESO 2.2-meter
telescope where the average galaxy number density is $n_{\rm
  g}=12~{\rm arcmin}^{-2}$ with the median seeing of $0.9$ arcsec. They
examined the galaxy over-density on the $R$-band image around the peak
($<2'$) and found that 65\% of the peaks accompany the galaxy concentrations. 
The rest of the peaks were most likely artifacts due to the noise
because these preferentially appear on the shallower imaging
data. When they raised the peak detection threshold up to 6, all the
peaks have optical counterparts. \citet{2007ApJ...669..714M} showed
that nearly 75\% of the peaks ($\nu>4.5$) have optical or X-ray
counterparts on a 2.2~deg$^2$ field observed by 8.2-meter Subaru
with Suprime-Cam under the seeing of $0.55$ arcsec ($n_{\rm g} = 46~{\rm
  arcmin}^{-2}$). Even if the detection threshold is decreased down
to 4, the purity was kept high, around 80\%. This demonstrates that the
contamination level can be indeed reduced using deeper and sharper
imaging data. 

Following these pilot studies, weak lensing mass map peaks have been
searched on much wider survey area such as 4-meter CFHT Legacy Survey 
(CFHTLS). \citet{2012ApJ...748...56S} presented the results from a 51~deg$^2$
area ($\sim 30$\% of CFHTLS-Wide Survey area, $n_{\rm g}=11.5~{\rm
  arcmin}^{-2}$ with the median seeing of $0.71$ arcsec).  They found 51
mass map peaks with $\nu>4.5$, 59\% of which have counterparts in the
optical cluster catalogs generated using photometric redshifts from
the same CFHTLS data set \citep{2009ApJ...706..571T}. 
\citet{2015MNRAS.450.2888L} also identified $\sim 40$ mass map peaks
with $\nu>4.5$ from $\sim 130$~deg$^2$ of CFHT Stripe 82 Survey.
There have also been attempts to constrain cosmological parameters 
from the number counts of peaks, although not necessarily using high
$\nu$ peaks
\citep{2015MNRAS.450.2888L,2015PhRvD..91f3507L,2015PASJ...67...34H,2016MNRAS.463.3653K}. 

Shear selected cluster catalog can uniquely address the abundance 
of clusters with anomalously high mass-to-light ratios. The
possibility of such dark halos is argued by e.g., \citet{weinberg02}. 
However, previous shear selected cluster searches mentioned above
did not successfully distinguish dim clusters from false positive
peaks caused by the noise or the line-of-sight projection of small
systems, mainly due to the lower detection thresholds for selecting
peaks from mass maps. The lack of follow-up data might also have
complicated the cross-correlation of peaks and galaxy concentrations.

This paper presents the first attempt to generate a useful shear selected 
cluster catalog where the contamination is mitigated by adopting a
higher peak detection threshold than what was adopted by previous works
and the catalog is validated by
comparing with optically selected groups/clusters based on the
multi-color deep Hyper Suprime-Cam (HSC) data. HSC realizes seeing
limited imaging on Maunakea over the entire 1.5~degree diameter field
of view \citep{miyazaki17}. Combined with the large aperture of the
Subaru telescope, the camera is an ideal facility for weak lensing 
surveys. Using the commissioning data of HSC over 2.3~deg$^2$, 
\citet{2015ApJ...807...22M} indeed showed that the angular resolution
of the weak lensing mass map is fine enough to collect un-contaminated
cluster samples over the wide redshift range, $0.2 < z < 0.7$. 
In this paper, we extend the previous work to construct a large sample
of shear selected clusters from the HSC Subaru Strategic Program (SSP)
survey data, which already covered more than 100~deg$^2$ with full five
broadband colors. 

This paper is organized as follows. In Section~\ref{sec:data}, we
describe the HSC-SSP data used for generating weak lensing mass maps.
The mass maps are presented in Section~\ref{sec:map}, and peaks in the
mass maps are identified in Section~\ref{sec:peak}. We discuss
properties of our sample of shear selected clusters in
Section~\ref{sec:discussion}, and give conclusion in
Section~\ref{sec:conclusion}. 
Unless explicitly stated, we adopt a WMAP9 cosmological model with 
$\Omega_{\rm M} = 0.287$, $\Omega_\Lambda=0.713$, $\sigma_8 = 0.820$, $H_0 =
100h\;{\rm km}\;{\rm s}^{-1}\;{\rm Mpc^{-1}}$ with $h = 0.693$
\citep{2013ApJS..208...19H}. 

\section{Imaging data set and the shape measurements}\label{sec:data}

The HSC-SSP is a legacy optical imaging survey consisting of three
layers; Wide, Deep and UltraDeep. HSC-Wide is designed to be a
competitive cosmological survey in which nearly ten times larger
survey field is covered with one magnitude deeper imaging compared
with the existing CFHTLS. HSC-Deep and -UltraDeep uniquely combine
narrow-band imaging with broad-band imaging to explore frontiers of
studies in high redshift objects and galaxy evolution. In total 300
nights were awarded for the HSC-SSP. The survey started on March 2014
and will continue for about six years. The details of the HSC-SSP
survey is given in \citet{aihara17}.  

In this paper, we adopt the HSC-Wide data from an internal data release
called S16A.  The integration time of a single exposure is 200~sec in
$i$-band. The survey area is covered by the camera's circular field of
view with dithered pointings. The dithering pattern is determined so
that any location on the sky is visited by at least six exposures for
$i$-band, yielding the total exposure time of 1200~sec. The HSC-Wide
$i$-band imaging is preferentially conducted under good seeings,
resulting in a median seeing measured over three years of $0.56$ arcsec FWHM
\citep{miyazaki17}.  

The survey fields are mostly located along the equator (S16A field
names XMM, GAMA09H, WIDE12H, GAMA15H, and VVDS) except the one
(S16A field name HECTOMAP). The total area of the HSC-Wide S16A data
used for this work amounts to $\sim 160$~deg$^2$. 

The data reduction is made with the HSC pipeline
\citep{bosch17}, which is based on  ``LSST-Stack'' \citep{lsst-stack2,
  lsst-stack}, a software suite being developed for the LSST
project. We adopt the re-Gaussianization method \citep{hirata03} for
measuring shapes of galaxies. Detail of the Point Spread Function
(PSF) corrections, shear measurements, and the results of null tests
are shown in \citet{mandelbaum17}. In this HSC S16A shape catalog, the
faint magnitude cut is conservatively chosen to $i<24.5$. With this
conservative cut, however, the number density of source galaxies is
still high, with a raw number density of $n_{\rm g}=25~{\rm
  arcmin}^{-2}$. Additional systematics tests using weak lensing mass
maps for the HSC S16A shape catalog are given in \citet{oguri17b}, in
which small residual systematics in weak lensing mass maps are
confirmed. While the mass maps are constructed without applying any
cut in photometric redshifts of source galaxies, we adopt {\tt mizuki}
photometric redshift \citep{tanaka17} to select background galaxies
behind clusters when we conduct radial profile fitting for 
individual shear selected clusters. We note that the selection of
photometric redshift in \citet{tanaka17} does not affect the
individual mass estimate compared with the error.  

\section{Weak lensing mass maps}\label{sec:map}

\subsection{Reconstruction of mass maps}\label{sec:mapmake}

In order to locate the position of dark matter concentrations by weak
lensing, we generate two-dimensional mass map from the shear
catalog. Since the reconstructed mass map is noisy, 
smoothing is crucial for detecting any structure in the map.
A useful statistics is the so-called aperture mass
\citep{1996MNRAS.283..837S}, which uses essentially a smoothed lensing
convergence field $\kappa(\boldsymbol{\theta})$ defined as  
\begin{equation}
  M_{\rm ap}(\boldsymbol{\theta}_0) = \int d^2\theta \; 
\kappa(\boldsymbol{\theta}) U(|\boldsymbol{\theta}-\boldsymbol{\theta}_0|),
\end{equation}
where $U(\theta)$ is a circular-symmetric filter function. When a
compensated filter is chosen, $\int d\theta \theta U(\theta) = 0$, the
zero point of kappa does not contribute to $M_{\rm ap}$.  Using the
shear $\gamma$, $M_{\rm ap}$ can be rewritten as 
\begin{equation}
  M_{\rm ap}(\boldsymbol{\theta}_0)  = \int d^2\theta \gamma_T(\boldsymbol{\theta},
  \boldsymbol{\theta}_0)\;Q(|\boldsymbol{\theta}-\boldsymbol{\theta}_0|),
  \label{mapgt}
\end{equation}
where $\gamma_T$ is the tangential shear at the point
$\boldsymbol{\theta}$ with respect to the point
$\boldsymbol{\theta}_0$, and $Q(\theta)$ is related to $U(\theta)$ as 
\begin{equation}
  Q(\theta) = \frac{2}{\theta^2}\int_0^\theta d\theta' \theta' U(\theta') -
  U(\theta). 
\end{equation}

The filter $U(\theta)$ must be chosen carefully depending on the
purpose. In order to construct a shear selected cluster sample
efficiently, the angular size of the filter has to be roughly the
scale radius of massive clusters of interest \citep{hamana04}. 
In order to maximize the signal from mass concentration with NFW
\citep{navarro97} profile, it is most efficient to choose the spatial
filter that follows the radial convergence profile of the NFW
profile \citep{2005A&A...442..851M,2006PhRvD..73l3525M}. However, near
the center there are many observational uncertainties, including the
non-linearity in the shape measurement, the dilution by cluster member
galaxies, and the large baryonic effect that modifies the density
profile near the center. Thus in practice we want to choose the filter
$Q(\theta)$ that suppresses the contribution from the cluster center
(i.e., small $|Q(\theta)|$ at $\theta\approx 0$). Also it is useful
to consider a compensated filter for $U(\theta)$  in order to
minimize the effect of the large-scale structure in selecting clusters
from mass map peaks \citep{2010ApJ...709..286M}.

In this paper, we consider a generic form for simplicity, adopting a
truncated Gaussian \citep{2012MNRAS.425.2287H}
\begin{eqnarray}
  U_{\rm G}(\theta) & = & \frac{1}{\pi\theta_s^2}\exp\left({-\frac{\theta^2}{\theta_s^2}}\right)-U_0\\
  Q_{\rm G}(\theta) & = &\frac{1}{\pi \theta^2} \left [1 - \left (1 + 
  \frac{\theta^2}{\theta_s}\right )\exp\left({-\frac{\theta^2}{\theta_s^2}}\right) \right ],
\end{eqnarray}
for $\theta \leq\theta_{\rm out}$ and $U_{\rm G}(\theta)=Q_{\rm
  G}(\theta)=0$ for $\theta>\theta_{\rm out}$. The parameter
$\theta_s$ represents the angular scale of the filter, whereas $U_0$
is introduced so as to satisfy the condition $\int d\theta \theta
U(\theta) = 0$. Then the aperture mass is calculated in the
two-dimensional grids over the observed data to yield the 
$M_{\rm ap}$ map. In practice the integral in equation~(\ref{mapgt})
is replaced by a discrete sum of tangential shears estimated from
image ellipticities 
\begin{equation}
  M_{\rm ap}(\boldsymbol{\theta}_0) = \frac{1}{\bar{W}(\boldsymbol{\theta}_0)}\sum_i w_i \epsilon_{T,i}(\boldsymbol{\theta}_i,
  \boldsymbol{\theta}_0)\;Q(|\boldsymbol{\theta}_i - \boldsymbol{\theta}_0|),
\end{equation}
\begin{equation}
 \bar{W}(\boldsymbol{\theta}_0)= \frac{1}{\pi
   \theta_{\rm out}^2}\sum_{|\boldsymbol{\theta}_i -
   \boldsymbol{\theta}_0|<\theta_{\rm out}} (1+m_i)w_i,
\end{equation}
where $\epsilon_{T,i}(\boldsymbol{\theta}_i, \boldsymbol{\theta}_0)$ 
is the tangential component of the
ellipticity of the $i$-th galaxy located at $\boldsymbol{\theta}_i$
with respect to the center of the aperture, $\boldsymbol{\theta}_0$. 
The parameter $w_i$ is the shear weight and $m_i$ is the multiplicative bias
for the $i$-th galaxy, both of which are derived in \citet{mandelbaum17}. 

The noise of $M_{\rm ap}(\boldsymbol{\theta}_0)$ is estimated from the
variance of the $M_{\rm ap}$ values where the galaxy orientations are
randomized in each realization. When we randomize the rotation angle
$\phi_i$ uniformly from 0 to 2$\pi$, the variance is calculated by
\begin{eqnarray}
  \sigma^2(\boldsymbol{\theta}_0) & = & \frac{1}{2\pi}  \int_0^{2\pi}
    d\phi_1\cdot\cdot\cdot d\phi_n M^2_{\rm ap}(\boldsymbol{\theta}_0, \phi_i) \nonumber\\ 
   & & -  
\frac{1}{2\pi}\left( \int_0^{2\pi}  d\phi_1\cdot\cdot\cdot d\phi_n M_{\rm ap}(\boldsymbol{\theta}_0, \phi_i)
  \right)^2  \\
 & = & \frac{1}{2\left\{\bar{W}(\boldsymbol{\theta}_0)\right\}^2}
\sum_i\left\{w_i|\epsilon_{T,i}|Q(|\boldsymbol{\theta}_i -
\boldsymbol{\theta}_0|)\right\}^2. 
\end{eqnarray}
We define the signal-to-noise ratio of $M_{\rm ap}$ at the point 
$\boldsymbol{\theta}_0$ as 
\begin{equation}
{\nu}(\boldsymbol{\theta}_0) = 
\frac{M_{\rm ap}(\boldsymbol{\theta}_0)}{\sigma(\boldsymbol{\theta}_0)}.
\label{defnu}
\end{equation}

Throughout the paper, we employ reduced shear estimated from the shape
measurements as an approximation of the shear. This makes $\nu$ slightly
increased; 2 \% and 5 \% for $\nu = 4.7$ and 9.0 peak, respectively.
We also note that the multiplicative bias $m$ does not affect the value of
$\nu$ for each grid because it changes both $M_{\rm ap}$ and $\sigma$
in the similar manner. However, it changes the absolute value of
$M_{\rm ap}$ and $\sigma$, and therefore changes the cluster mass
scale for a given $\nu$. 

\begin{figure*}
 \begin{center}
  \includegraphics[width=17cm]{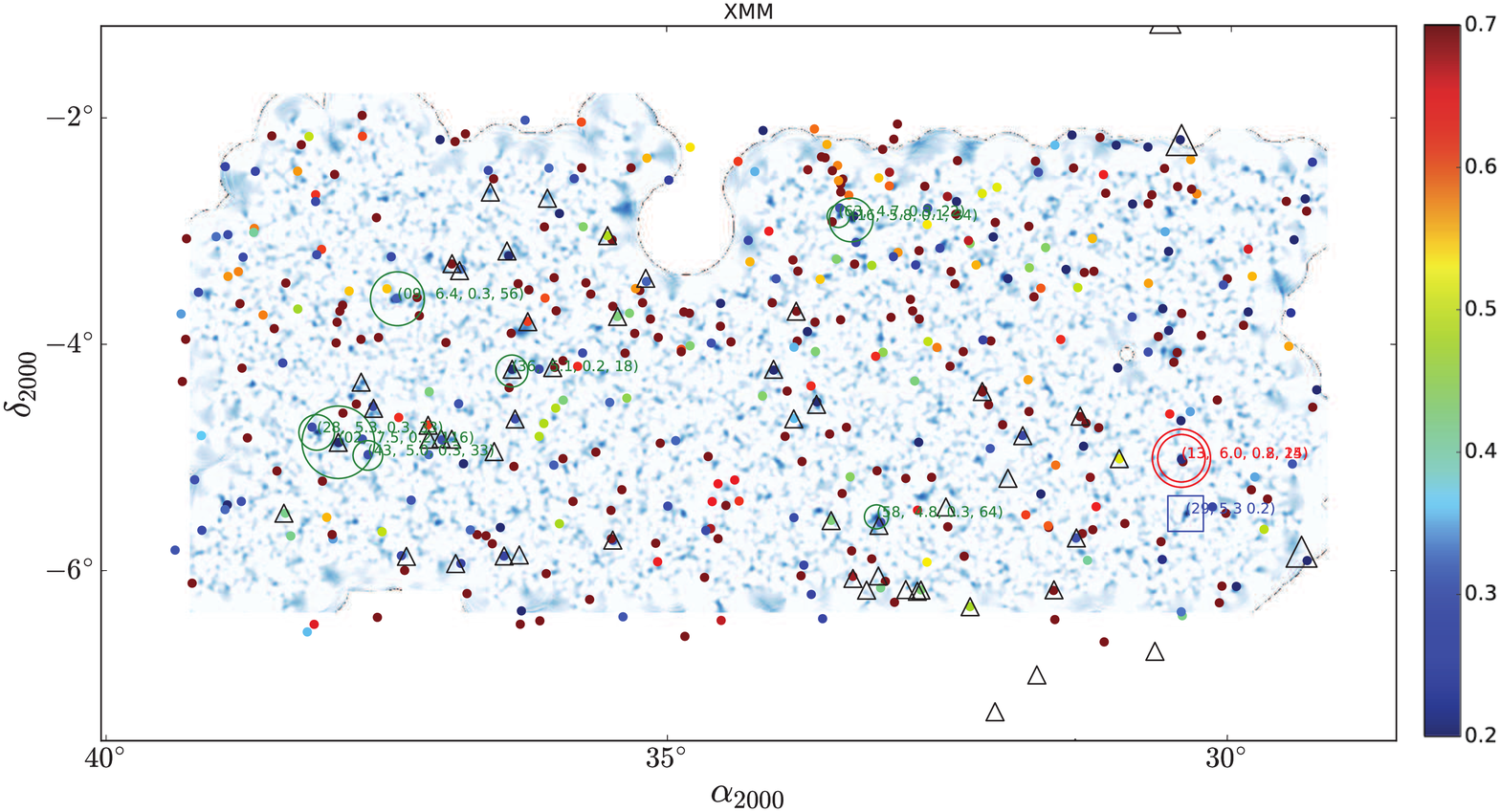}
  \includegraphics[width=17cm]{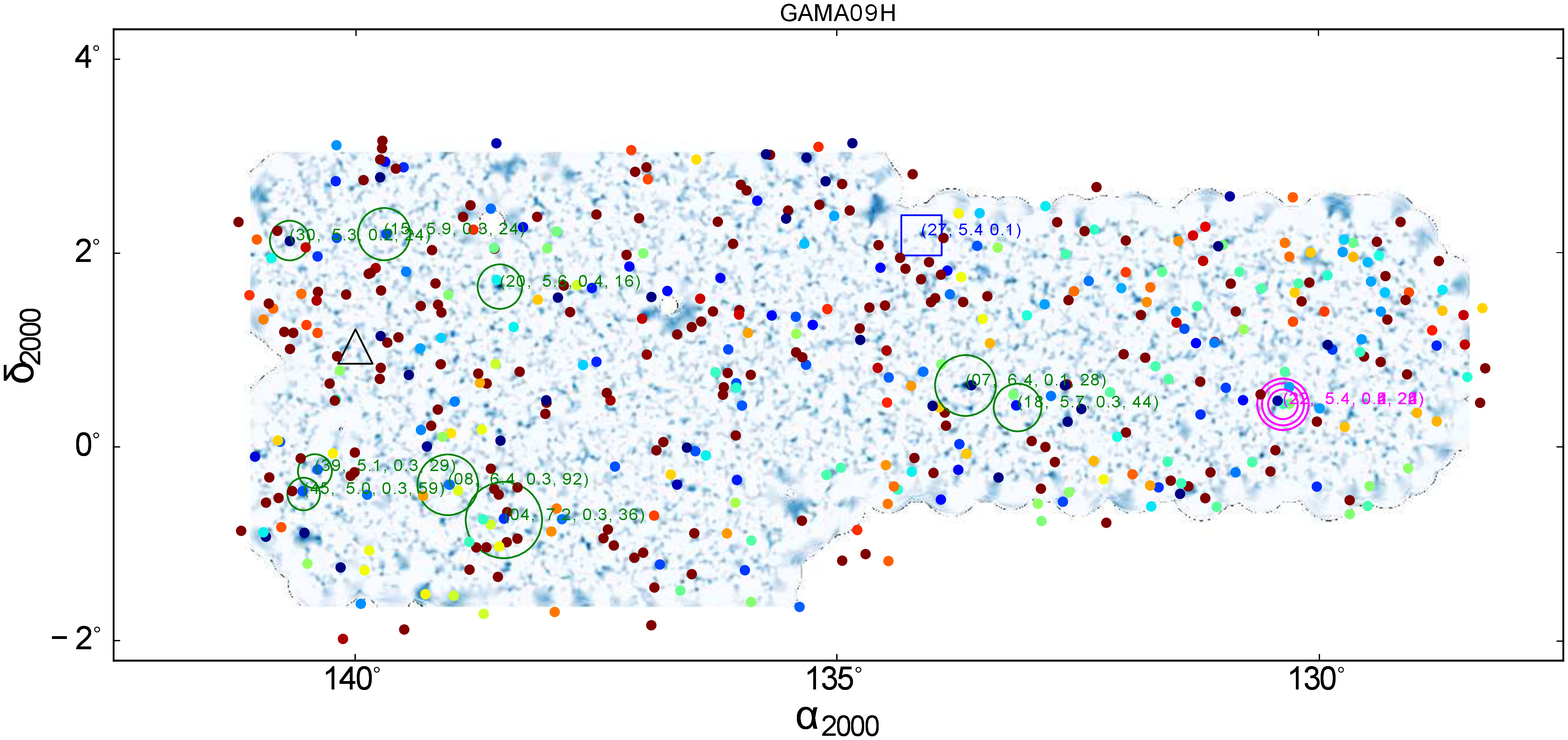}
 \end{center}
 \caption{Weak lensing $M_{\rm ap}$ SN maps for the XMM ({\it upper})
   and GAMA09H ({\it lower}) fields. Small filled circles show the locations
 of CAMIRA clusters where blue to red color-code corresponds the redshift of 0.2
 to 0.7 as is shown in the color bar next to the map of XMM field. Open circles
 show the locations of the mass map  peaks matched with CAMIRA clusters. If
 multiple CAMIRA clusters are matched
 with a peak, the number of concentric circle indicate the multiplicity. 
 The first two numbers in brackets show the peak rank and the peak SN. The third
 numbers and the forth number show the redshift and richness, respectively, but
 these are not always available. Peaks that do not match with CAMIRA clusters are
 shown by open squares. Large open triangles indicate MCXC X-ray clusters. Small
 open triangles show the XXL brightest 100
 clusters. Large open diamonds show the location of the peaks that matches
 neither CAMIRA nor WHL15 clusters.} \label{mapmap1}
\end{figure*}

\begin{figure*}
 \begin{center}
  \includegraphics[width=17cm]{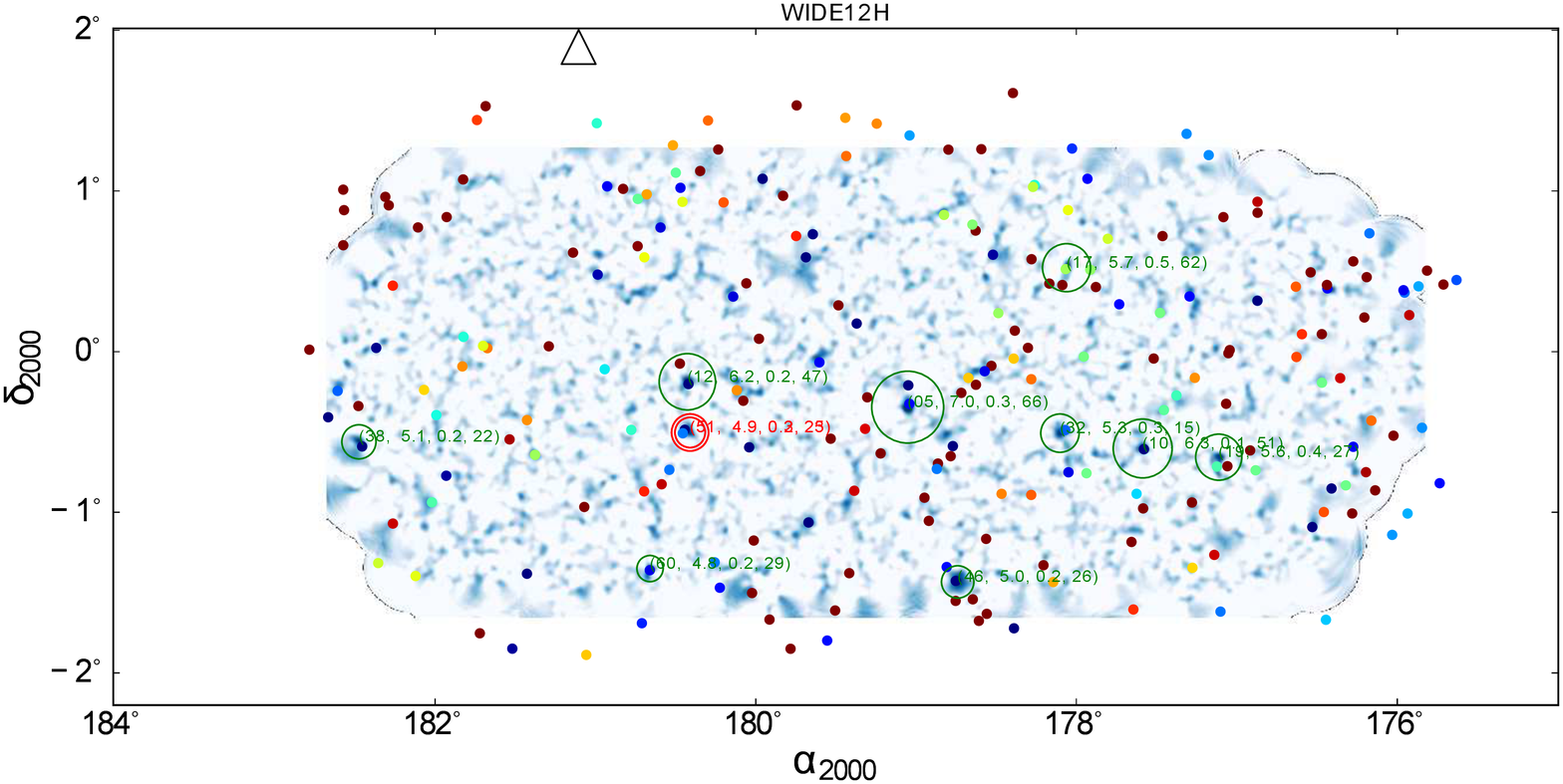}
  \includegraphics[width=17cm]{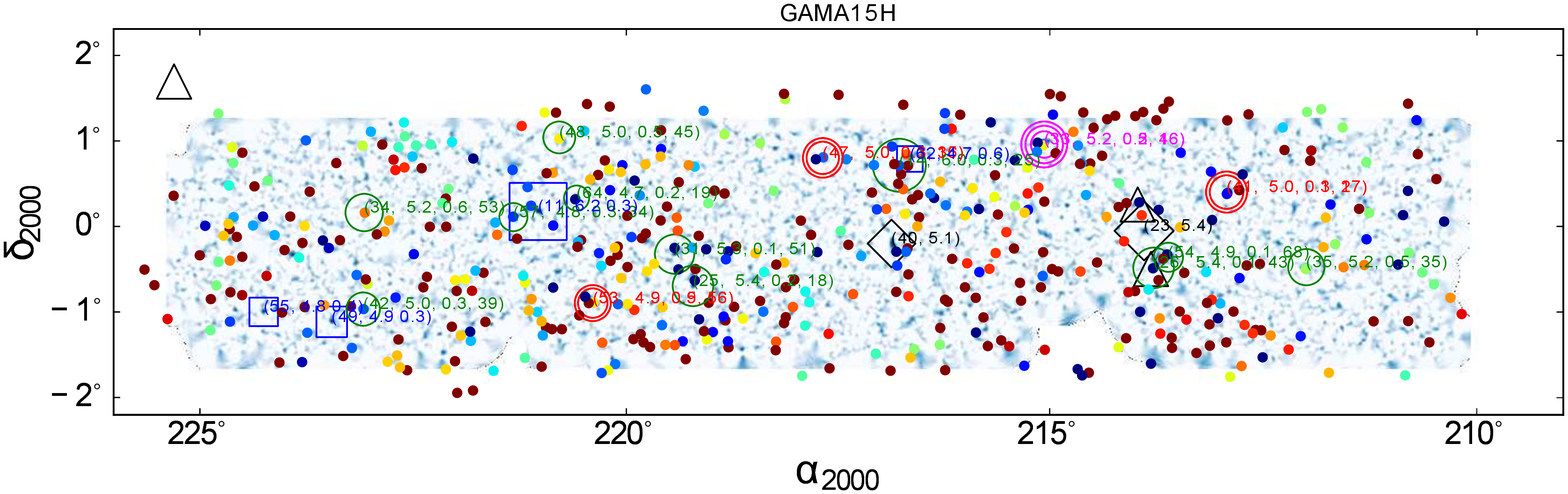}
 \end{center}
 \caption{Same as Figure~\ref{mapmap1}, but weak lensing $M_{\rm ap}$
   SN maps of WIDE12H ({\it upper}) and GAMA15H ({\it lower}) are
 shown.} \label{mapmap2}
\end{figure*}

\begin{figure*}
 \begin{center}
  \includegraphics[width=17cm]{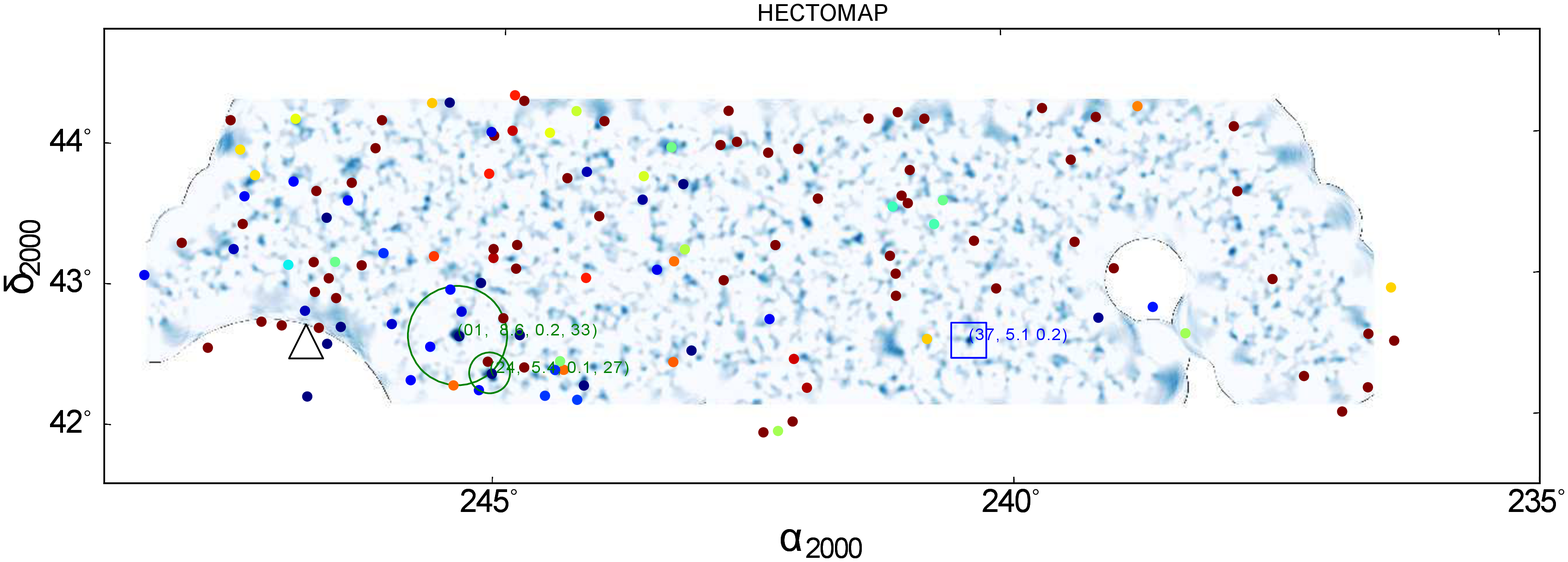}
  \includegraphics[width=17cm]{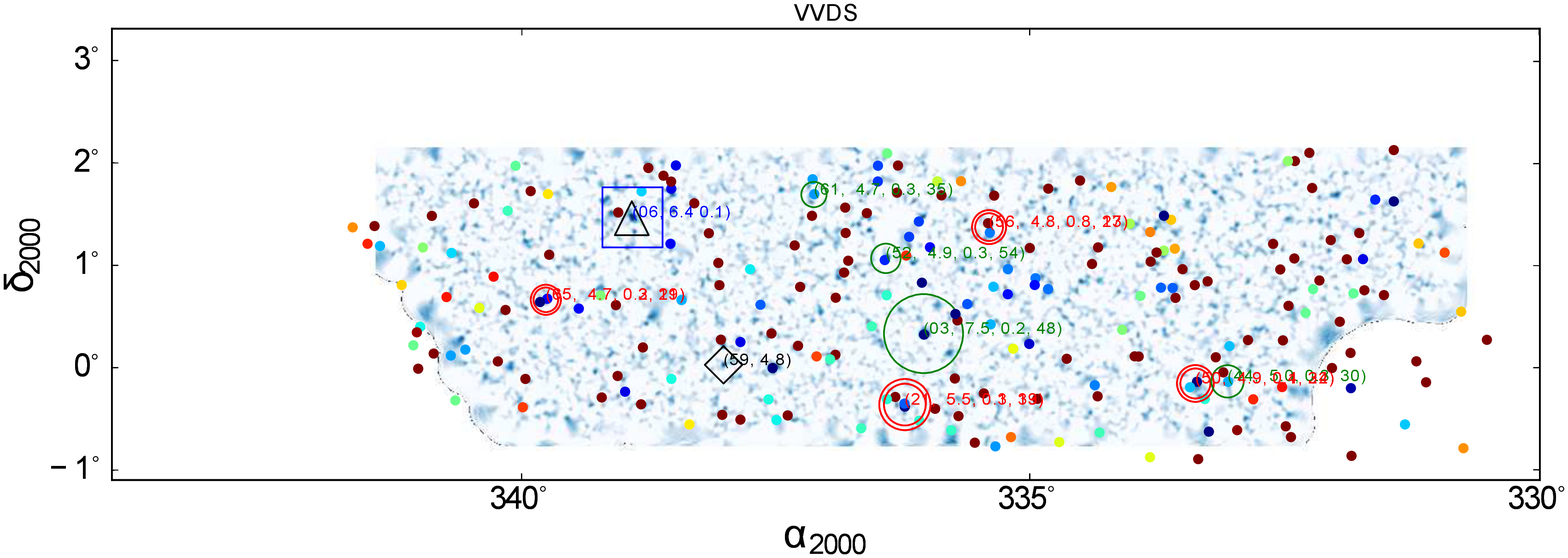}
 \end{center}
 \caption{Same as Figure~\ref{mapmap1}, but weak lensing $M_{\rm ap}$
   SN maps of HECTOMAP ({\it upper}) and VVDS ({\it lower}) are shown.
   }
 \label{mapmap3}
\end{figure*}

Figures~\ref{mapmap1}, \ref{mapmap2}, and \ref{mapmap3} show the SN
map of each field when 
we adopt the truncated Gaussian filters of $\theta_s = 1.5'$ with the
map grid spacing of $0.25'$. The filter is truncated at $\theta_{\rm
  out} = 15'$. 
Objects in the shear catalog have various flags such as a bright
  star flag \citep{mandelbaum17}.
We eliminate objects with such flags from lensing analysis and this acts as
masking of inappropriate field position.

The peaks on the SN map indicates the locations of the
mass concentrations of dark matter halos. We recognize the local peaks
when three connected pixels on the map exceeds the threshold of $\nu
= 4.3$. The location of the peak is simply read from the center of the
grid whose $\nu$ is maximum.  The maximum value is assigned for the SN
of that peak.  

Based on the discussion made in Section~\ref{bmode}, we set 
the peak SN threshold at $\nu=4.7$ in this paper. The number of peaks
depends on the choice of the filter scale $\theta_s$. We adopted
the value that is slightly larger than \citet{2012MNRAS.425.2287H},
who adopted $\theta_s = 1'$, because in this paper we only consider
highly significant peaks whose counterpart clusters tend to be located at
relatively low redshift and have larger angular extents. Indeed, the
numbers of peaks with the two different filters are 56 and 65 for
$\theta_s=1'$ and $1.5'$, respectively. In principle, we can consider
multiple mass maps with different filter sizes, which will definitely
increase the information content extracted from the mass maps 
\citep{2012MNRAS.423.1711M,2016A&A...593A..88L}, but in this paper we
fix the filter size throughout for simplicity.

\subsection{Peaks on the $B$-mode map and the rate of false signals} \label{bmode}

\begin{figure*}
 \begin{center}
  \includegraphics[width=17cm]{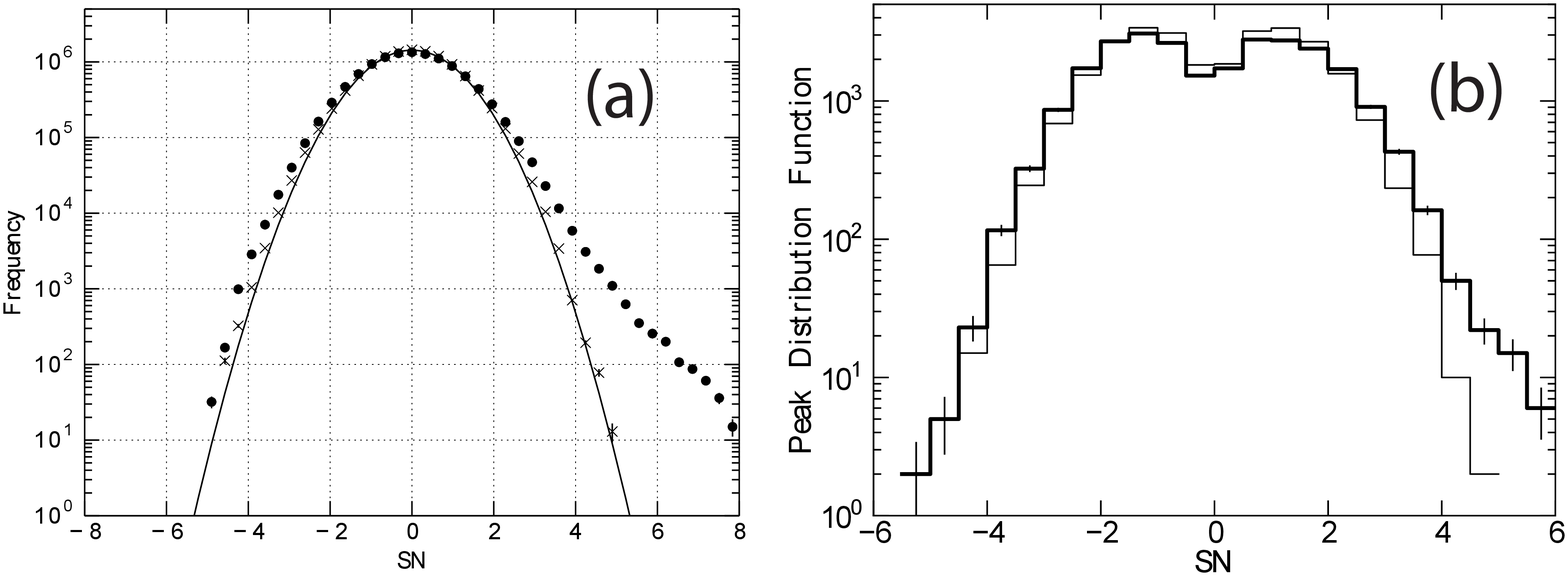}
 \end{center}
 \caption{(a) Pixel distributions of the $E$-mode ($\bullet$) and
   $B$-mode ($\times$) $M_{\rm ap}$ SN maps. Error bars of the data are shown
   which are only visible at $|\rm{SN}| > 4$. 
$B$-mode distribution  nicely follows the Gaussian distribution with
   $\sigma=1$. (b) Local peak distribution function of the same maps. Local
   peaks on the E-mode  maps is shown by thick histogram whereas peaks on B-mode
   maps by thin  histogram.}\label{wlpdf}
\end{figure*}

Because the weak lensing is induced by the scalar (gravitational) potential, no
$B$-mode signal is generated from lensing in principle. We can use this fact to
see whether we have systematic errors in our analysis. $B$-mode SN map can be
made from the same galaxy shape catalog where the each galaxy
orientation is rotated by 45 degree. In Figure~\ref{wlpdf} (a), we show
the pixel value distribution function of the $B$-mode map as cross
symbols ($\times$).  As is shown, there is no pixel value beyond
$\nu=5$. Solid line shows the Gaussian distribution with $\sigma=1$
with the amplitude normalized to the $B$-mode pixel frequency at $\nu
= 0$, which nicely agrees with the distribution of the observed
$B$-mode distribution. This suggests that $B$-mode map is a Gaussian
random field and is not significantly affected by systematic errors. 
This result, together with the analyses presented in
\citet{mandelbaum17} and \citet{oguri17b}, suggests that systematic
errors on the $E$-mode mass maps are also not significant and highly
significant peaks ($\nu > 5$) on the $E$-mode maps are not likely to  
be generated by the noise or systematic errors. 
The numbers of less significant peaks on $E$-mode and $B$-mode map
are 22 and 4, respectively. Therefore, up to $\sim 20$\% of such less
significant $E$-mode peaks can be false signals caused by the noise.  

Figure~\ref{wlpdf}~(b) shows a local peak height (and trough depth for negative
SN) distribution function. Here the local peak (trough) is identified by  pixels on
the maps whose pixel value are higher (lower) than any values in the surrounding
eight pixels  This is another representation of the nature of the maps and can
be directly compared with the theoretical calculation \citep{jain00}. We see
significant excess of the peak counts over noise estimated by B-mode peak
counts. We could also see the small excess of trough over the noise at SN $<$ 4.5.

\section{Identifications of the peaks}\label{sec:peak}

\begin{longtable}{llrrllllll}
\caption{List of shear selected clusters} \label{papertablephotoz} \\
\hline
rank & SN ($\nu$) & RA2000 & DEC2000 & d        & $z_{\rm cl}$ & $N_{\rm mem}$ & $M_{500}$& $c_{500}$  & Note \\
     &    & \multicolumn{1}{l}{[deg]}  & \multicolumn{1}{l}{[deg]}   & [$h^{-1}$Mpc] &          &          & $\times 10^{14} h^{-1}$M$_\odot$&  &\\ \hline \endhead
\hline
\input papertablephotozmodclimit.tex
\hline
\end{longtable}

In total we identify 65 peaks with $\nu>4.7$ for the smoothing size
$\theta_s=1.5'$, which are summarized in Table~\ref{papertablephotoz}. 
The peaks are sorted by $\nu$ and we call the order 'rank'. 
We now compare the peaks with optical and X-ray cluster catalogs. 

\subsection{Optical counterparts} \label{opticalcounterparts}

Based on the same HSC-SSP multi-color photometric data used in this paper,
\citet{oguri17a} generated an optical cluster catalog using the ``Cluster finding
algorithm based on Multi-band Identification of Red-sequence gAlaxies'' (CAMIRA),
which was developed in \citet{oguri14}. CAMIRA makes use of the
stellar population synthesis (SPS) model of \citet{bruzual03} to
compute SEDs of red-sequence galaxies, estimates the likelihood of them
being cluster member galaxies for each redshift using $\chi^2$ of the
SED fitting, constructs a three-dimensional richness map using a
compensated spatial filter, and identifies cluster candidates from
peaks of the richness map. The richness threshold is set to $N_{\rm
  mem}=15$ in identifying the clusters.  

The locations of the CAMIRA clusters are overlaid on Figures~\ref{mapmap1},
\ref{mapmap2}, and \ref{mapmap3}, as small
filled circles. Clusters are searched around the peaks with the a
loose matching tolerance of $6'$ at first. Next, we calculated the
comoving distance between the peak and the cluster center using the
estimated cluster redshift. The cluster center
is the position of the brightest cluster galaxy, BCG,  recognized through the
CAMIRA algorithm.
Then, we identify the peak with the
cluster when the distance is within $1.5h^{-1}$Mpc. In some cases,
multiple clusters are matched with a peak where cares must be taken
because the weak lensing mass estimate is complicated for such peaks.

In Table~\ref{papertablephotoz}, we list the coordinates of the detected peaks
sorted by the SN together with the CAMIRA redshift, $z_{\rm cl}$, the
richness, $N_{\rm mem}$, and the distance, $d$, between the peak and
the cluster location. Open circles on Figures~\ref{mapmap1},
\ref{mapmap2}, and \ref{mapmap3} show the locations of the peaks
matched with a single CAMIRA cluster. When multiple clusters are
matched, the multiplicity is indicated by the number of concentric
open circles. The diameter of the circle roughly reflect the rank of
the cluster such that highly ranked peaks have larger
circles. Table~\ref{matchwithcamira} summarizes the multiplicity of
the matching.

\begin{figure}
 \begin{center}
  \includegraphics[width=8cm]{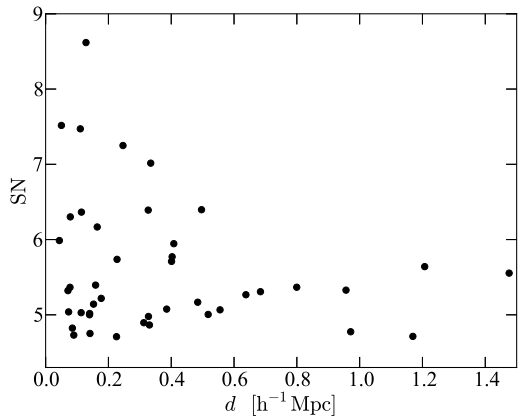}
 \end{center}
 \caption{
The distance between the peak position and the cluster center (BCG position) of
CAMIRA clusters. 
}
   \label{d_sn_39}
\end{figure}

The angular resolution of the weak lensing mass map is usually poorer than
that of X-ray map and therefore, the peak position could
not be the best proxy for the real dark matter halo center.
However, the offset between the real center and our detected peak cannot largely
exceed 1.5 arcmin because we adopted the smoothing scale of 1.5 arcmin.
The simulation study done by \citet{dietrich12} also shows
that the offset between the peak position and dark matter halo center is usually smaller
than 1 arcmin. On the other hand, the cluster center (the BCG position) is known
to have larger scatter up to $1h^{-1}$Mpc with respect to the X-ray center as
is shown in Fig.12 of \citet{oguri17a}. Therefore, in this work, we adopt the
peak position as an approximate position of dark matter halo center because not all of X-ray
data are in hand.
Figure~\ref{d_sn_39} shows the distance between the peak and the cluster center
versus SN of the peak where the larger scatters is observed for less significant
peaks. This might suggest that the physical association of the peak with the
clusters is unlikely whose separation is as large as $1.5h^{-1}$ Mpc although we
are not certain how large the actual scatter of BCG position is in CAMIRA
catalog.

We have 11 peaks which have no counterpart in the CAMIRA cluster
catalog. What is the nature of these peaks? We search for the counterparts in
a cluster catalogs generated by \citet{whl15} (WHL15), which is based
on the Sloan Digital Sky Survey, with the same matching tolerance of
$1.5h^{-1}$Mpc in physical distance. Eight peaks out of the 11 peaks
are found to have the counterparts within $0.5h^{-1}$Mpc. The mean
richness of the entire WHL15 clusters is 24.7. Half of the matched
counterparts have richness that exceed the mean and the other half
have the richness below the mean. 

Each cluster catalog has different selection criteria. For example,
CAMIRA has set the lowest redshift limit to $z=0.1$ and this explains
why the peak rank 6 (Abell 2457 at $z=0.0594$) is unmatched with
CAMIRA catalog. The richness threshold is also different such that
CAMIRA richness threshold is more conservative than WHL15. We note
that the definitions of the richness are different between CAMIRA and
WHL15, and therefore we cannot directly compare the richness values
between these two catalogs. 

\begin{table}
\caption{Optical identifications of the peaks}
\vskip 3mm
\begin{tabular}{llll} \hline
\multicolumn{2}{c}{CAMIRA} & \multicolumn{2}{c}{WHL15}  \\
N$_{\rm match}$ & N$_{\rm peak}$ & N$_{\rm match}$ & N$_{\rm peak}$\\ \hline
1 & 43 & & \\
2 & 9 & & \\
3 & 2 & &\\
0 & 11 & 1 & 8\\
  &   &  0 & 3 \\ \hline
\end{tabular}
\label{matchwithcamira}
\begin{tabnote}
Number of clusters matched with the peaks under the tolerance of $1.5h^{-1}$Mpc.
\end{tabnote}
\end{table}

Three peaks are still unmatched with neither with CAMIRA nor WHL15 (rank = 23,
40, 59). We inspect the images around peak 23, and find, with a distance of
$21.3$ arcsec from the peak position, a peculiar group of galaxies that
consist of five cores. The system was studied by \citet{filho14} in
which they confirmed that it is indeed a multiple dry merger at $z =
0.18$. It is however unlikely for this system alone to generate the
observed lensing signal of $\nu=5.42$ unless we assume the unusually
high mass-to-light ratio. We revisit the three-dimensional richness
map where the CAMIRA clusters are searched, and find that there is a
low richness ($N_{\rm mem}\sim10$) group at $z=0.18$ which coincides
with the peak 23. It is also found that the galaxy catalog, on which
the CAMIRA cluster was based, does not include these five galaxy
cores. It appears that the existence of a bright star ($B=14$) close
to the mergers (distance of $16.6$ arcsec) masks out these galaxies from
the galaxy catalog. Once we count on these five galaxies that undergo
dry merger, the richness of this cluster candidate reaches the
richness threshold ($N_{\rm mem}=15$) set for CAMIRA cluster search. 
Therefore, we conclude that the lensing signal is likely to be
generated by this galaxy cluster at $z=0.18$. 

As for the peaks 40 and 59, their $\nu$ do not exceed five
significantly, implying that these two peaks can be induced either
by statistical errors or a chance projection of small systems along
the line-of-sight. 
Following \citet{utsumi14}, we evaluate a probability to find a
spurious peak with $\nu \sim 5$ in the weak lensing map analytically. Given a
smoothing length of $\theta_s = 1.5'$ and an effective area of
160~${\rm deg}^2$, the analytical calculation gives the probability to find a
spurious peak with $\nu \sim 5$ is $\sim 2 \sigma$ level.
Also, we calculate an expected number of spurious peaks exceeding $\nu=4.7$ on
the 160~${\rm deg}^2$ weak lensing map using Equation 10 in
\citet{utsumi14}. The resulting expected number is 0.17 peaks per 160~${\rm
  deg}^2$. This number lies on the range of probabilities of 0.9772 
($2\sigma$) and 0.99 ($2.33\sigma$) if we quote the Poisson
single-sided lower limits for a value of 2 \citep{gehrels86}. Those calculations
show the $\nu$ value of the highest spurious peak and the number of spurious
peaks above the threshold are slightly larger than the case for the completely
ideal map but still in the range of $\sim 2\;\sigma$ level. Thus those
two spurious peaks can be explained by statistical errors.
In summary, out of 65 peaks with $\nu > 4.7 $ over
$\sim 160$~deg$^2$, we find 63 peaks have the physical
counterparts. Two peaks ($\nu=5.06$, $4.77$) can be false positives.\footnote{
We note that peak 40 has a counterpart with a separation of 30 arcsec in a
catalog by \citet{goto02} (SDSS CE J216.867157-00.209108) $z=0.174$ and peak 59
matches with a group at $z=0.03855$ identified by \citet{berlind06}. Because their
clusters are more abundant due to their choice of lower detection threshold, the
match could be accidental.
}

In Figure~\ref{wlpdf} (a) we find that the negative $E$-mode signal is
visible above the noise level. This should reflect the existence of
less dense region (chain of voids) along the line-of-sight. We
identify two significant trough of $\nu < -5$ at (RA2000, DEC2000) =
($216.947$, $-0.234896$) and  ($179.606$, $0.312002$). Visual
inspections of these regions indeed show that they are less populated
by galaxies. 

\subsection{X-ray counterparts} \label{xraycounterpart}

\begin{figure*}
 \begin{center}
  \includegraphics[width=15cm]{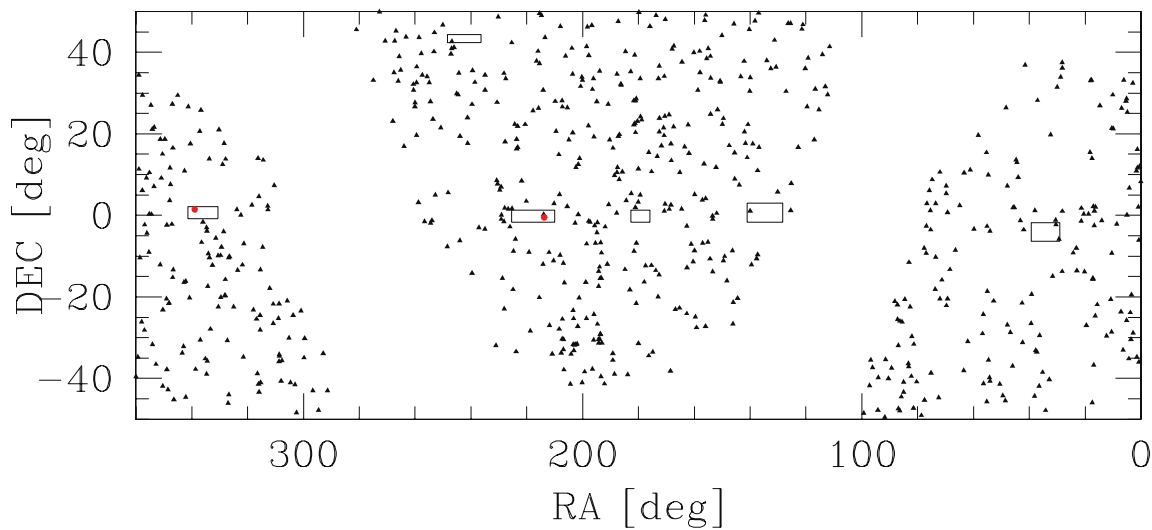}
 \end{center}
 \caption{Sky distribution of NORAS/REFLEX clusters compiled in the MCXC catalog
({\it solid triangles}) and the survey area of HSC-Wide S16A analyzed in this paper ({\it
     rectangles}). The MCXC clusters that are matched with the
   peaks are shown in red filled circles. There are four clusters in our survey
   regions, and among them, two are matched with the peaks.} 
 \label{skymap}
\end{figure*}

We now correlate our mass map peaks with published X-ray clusters to
compare the selection functions of lensing and X-ray cluster searches. 
\citet{2011A&A...534A.109P} sorted out catalogs of X-ray clusters to
generate a useful meta-catalog, called MCXC. Among them, we are
interested in the catalogs based on the ROSAT All Sky Surveys (RASS)
composed of the northern NORAS and southern REFLEX 
\citep{1999A&A...349..389V}. Figure~\ref{skymap} 
shows the sky distribution of the NORAS/REFLEX clusters compiled in the MCXC
catalog. The Figure indicates that our survey fields are buried well within the
area where the X-ray clusters were searched. We find that two peaks are matched
with the X-ray clusters under the tolerance of $5'$ (see
Table~\ref{matchmcxc}).  Here we adopt the slightly increased
tolerance radius of $5'$  following the argument by
\citet{2012ApJ...748...56S} where they argue the existence of additional noise
in the X-ray centers. 
The Table also shows that there are two
X-ray clusters which are within our survey field but have no
counterpart in the peak list. The redshifts are 0.0175 and 0.1259, respectively,
and the latter has a counter part in the CAMIRA catalog although the former does
not because of its low redshift value (the CAMIRA cluster catalog is
constructed only at $z>0.1$). The locations of these four X-ray
clusters are shown by large open triangles in Figures~\ref{mapmap1},
\ref{mapmap2}, and \ref{mapmap3}. 

\begin{table*}
\caption{Comparison of the peaks with the MCXC X-ray cluster catalog}
\vskip 2mm
\begin{tabular}{lllllll} \hline
rank & $z^{(1)}$ & $M_{\rm 500, WL}$ & $M_{\rm 500(MCXC)}$  & $L_{\rm 500,(MCXC)}$ &
Abell & MCXC \\
     &     & [$\times 10^{14}M_\odot$] & [$\times 10^{14}M_\odot$] &[$\times
  10^{44}{\rm erg\,s^{-1}}$] & & \\ \hline
6    & 0.0594 &  2.5$\pm$0.6  & 1.8   & 0.88  & 2457 & J2235.6+0128 \\
26   & 0.1403 &  2.8$\pm$0.9  & 2.8   & 1.9   & 1882 & J1415.2-0030 \\
-    & 0.0175 &   $< 0.5$  (2$\sigma$)  & 0.13  & 0.011 & - & J0920.0+0102 \\
-    & 0.1259 &   $< 0.6$  (2$\sigma$)  & 1.2   & 0.47  & - & J1415.8+0015 \\ \hline

\end{tabular}
\label{matchmcxc}
\vskip 1mm
\begin{tabnote}
Two MCXC clusters are matched with the peaks under the tolerance of
$5'$. The last two X-ray clusters do not have the counterparts in our
peak list. The redshifts $z^{(1)}$ are taken from the MCXC
catalog. $L_{\rm 500,(MCXC)}$ is measured in the 0.1 -- 2.4~keV band.
\end{tabnote}
\end{table*}

One of our survey field is designed to overlap with the wide field survey by
the {\it XMM-Newton} satellite called XXL \citep{xxl1}. We compare the peak
list with X-ray clusters published in \citet{xxl1}.
The location of
the XXL brightest 100 clusters are presented as small open triangles
in Figure~\ref{mapmap1}. In the overlap region, we find 39 XXL clusters
among which three weak lensing peaks are matched under the tolerance
of $5'$, as is summarized in Table~\ref{matchxxl}. Turning the problem
around, two peaks do not have any counterparts in the cluster catalog
although they are well inside the XXL survey footprint. 

\begin{table*}
\caption{Comparison of the peaks with the XXL brightest 100 X-ray cluster catalog}
\vskip 2mm
\begin{tabular}{lllllll} \hline
rank & $z$ & $M_{\rm 500, WL}$ & XLSSC & $M_{\rm 500, MT(XXL)}$  & $L_{\rm
  500 (XXL)}$  & $M_{\rm 500, WL(XXL)}$ \\
     &   & [$\times 10^{14}M_\odot$] & & [$\times 10^{14}M_\odot$]&[$\times 10^{44}{\rm erg\,s^{-1}}$] & [$\times 10^{14}M_\odot$]\\ \hline
2    &  0.186 &  3.9 $\pm$ 0.7 &  091  & 5.1 $\pm$ 2.2  &  1.31 $\pm$ 0.02& 6.2$^{+2.1}_{-1.8}$\\
36   &  0.155 &  1.7 $\pm$ 0.5 &  041  & 1.0 $\pm$ 0.4  &  0.12 $\pm$ 0.007& 0.7$^{+0.6}_{-0.5}$\\
58   &  0.287 &  2.6 $\pm$ 1.1 &  111  & 4.0 $\pm$ 1.8  &  0.67 $\pm$ 0.03& 6.3$\pm$1.8\\ 
28    &  0.276 & 2.4 $\pm$ 0.7 &  -  & -  &  -  & -\\ 
43    &  0.272 & 2.5 $\pm$ 0.9 &  -  & -  &  -  & -\\ 
\hline
\end{tabular}
\label{matchxxl}
\vskip 1mm
\begin{tabnote}
Three XXL clusters are matched with the peaks under the tolerance of
$3'$. The last two peaks do not have the counterparts in the brightest 100 XXL
clusters. The redshifts $z$ are taken from the CAMIRA catalog. 
$M_{\rm 500, XXL-T}$ are taken from \citet{xxl2}, and $M_{\rm 500,
  XXL-WL}$ from \citet{xxl4}. $L_{\rm 500 (XXL)}$ is measured in the
0.5 -- 2~keV band. 
\end{tabnote}
\end{table*}

\subsection{Mass estimate of the clusters}\label{sec:tanshear}

For each peak, we estimate the cluster mass from the observed
tangential shear radial profile which is azimuthually averaged in the
annulus. We assume that the density of the dark matter halos follows
the NFW profile
\begin{equation}
  \rho(r) = \frac{\rho_s}{\left(r/r_s\right)\left(1 + r/r_s\right)^2},
\end{equation}
where $\rho_s$ is the characteristic density and $r_s$ is the scale
radius. The two-dimensional surface mass density is 
computed as
\begin{equation}
  \Sigma(R) = 2\int_0^{\infty}\rho(R, y)dy, 
\end{equation}
where $R$ is the projected radial distance. 
The differential surface mass density is then given by,
\begin{equation}
  \Delta\Sigma(R) \equiv \bar{\Sigma}(<R) - \Sigma(R),
  \label{diffsigma}
\end{equation}
and 
\begin{eqnarray}
  \bar{\Sigma}(<R)  =  \frac{2}{R^2}\int_0^{\infty}\Sigma(R')R' dR' = 
r_s\rho_s g(R, r_s), \label{barsigma}
\end{eqnarray}
where $g(R, r_s)$ is given by equation~(8) in \citet{ford16}. 

In this paper, we define the cluster radius $r_\Delta$ within which
the mass density is a factor of $\Delta$ larger than the critical mass
density, $\rho_c$. The mass inside the $r_\Delta$ is naturally
calculated as  
\begin{equation}
M_\Delta = \frac{4}{3}\pi r_\Delta^3 \rho_c \Delta. 
\label{massdelta}
\end{equation}
Here we introduce the concentration parameter $c_\Delta$ as
\begin{equation}
  c_\Delta = \frac{r_\Delta}{r_s}
  \label{cdelta}
\end{equation}
In the mean time, \citet{navarro97} showed that $\rho_s$ is given by 
\begin{equation}
\rho_s = \rho_c\frac{\Delta c^3_\Delta}{3\left[\ln{(1 + c_\Delta)} -
    c_\Delta/(1+c_\Delta)\right]},
\label{rhos}
\end{equation}
where we adopt $\Delta = 500$ in this paper.

Observationally, the differential surface mass density,
$\Delta\Sigma(R)$, is constrained by the tangential shear
$\gamma_T(R)$. In the case of spherically symmetric dark matter halo
which we assume throughout this paper,  
\begin{equation}
  \Delta\Sigma(R) = \Sigma_{\rm cr}\gamma_T(R), \label{diffsigmaobs}
  \label{deltasigmaR}
\end{equation}
where 
\begin{equation}
  \Sigma_{\rm cr} = \frac{c^2}{4\pi G D_l}\left <
  \frac{D_{ls}}{D_s}\right >^{-1},
 \label{sigmacrit}
\end{equation}
with $D_l$, $D_{ls}$, $D_s$ being the angular diameter distances to
the lens, between the lens and source galaxies and to the source,
respectively. The distance factor, $D_{ls}/D_s$ is averaged over the
background source galaxies whose photometric redshift is 5\% larger
than the redshift of the lens.  

By fitting the radial profile of the tangential profile, we can constrain
($\rho_s$, $r_s$) using equations~(\ref{diffsigma}), (\ref{barsigma}), and
(\ref{diffsigmaobs}), and then ($M_\Delta$, $c_\Delta$) is obtained using
equations~(\ref{massdelta}) to (\ref{rhos}). We assume a flat cosmology
where $\rho_c(z)$ is given by
\begin{equation}
  \rho_c(z) = \rho_{c}(z=0)\left[\Omega_{M0}(1+z)^3 + \Omega_{\Lambda 0}\right].
\end{equation}
Fitting is made in the physical scale between 0.2 to $7h^{-1}$Mpc for
all cases. 

\begin{figure}
 \begin{center}
  \includegraphics[width=8cm]{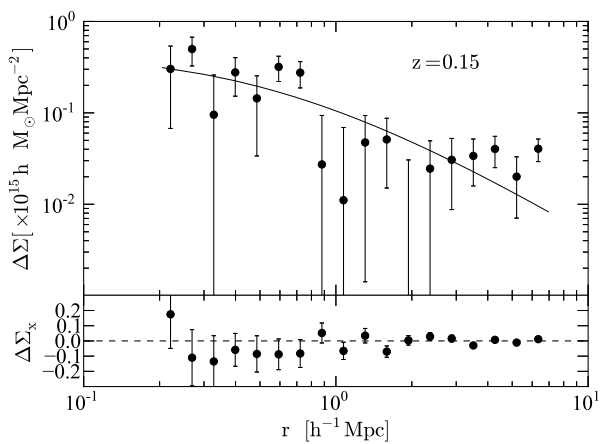}
 \end{center}
 \caption{Radial profile of the differential surface mass density of rank 4
   peak. The best-fit NFW model to the data is shown by the solid line. Bottom
   panel shows the profile generated by B-mode components.}
 \label{rdsigmarank4}
\end{figure}

As an example, Figure~\ref{rdsigmarank4} shows the observed radial
profile of differential surface mass density for rank 4 peak. The
error of each point is estimated from the orthogonal shear component. 
The best-fit NFW profile is shown by the solid line, which have the
mass and the concentration parameter values of 
($M_{500}$, $c_{500}$) = ($(2.97\pm 0.69)\times 10^{14}h^{-1}M_{\odot}$,
$2.1\pm1.0$). We present the ($M_{500}$, $c_{500}$) of all isolated
peaks (that matched with one CAMIRA cluster) in
Table~\ref{papertablephotoz}. The mass of multiply matched peaks are
not shown to avoid confusion. We note that the constraint on the
concentration parameter is poor for these individual peak analysis
because the constraint in the inner region is limited.

The measured mass depends largely on the outskirts of the dark matter halo and
we do not include the central part of the shears for fitting in order to avoid
the influence from dilution effect (See.~\ref{avradialprofile}), the error caused
by the uncertainties of the dark matter halo center is negligible.

\section{Discussions}\label{sec:discussion}

\subsection{Number of the observed peaks} \label{peakcount}

\begin{figure}
 \begin{center}
  \includegraphics[width=8cm]{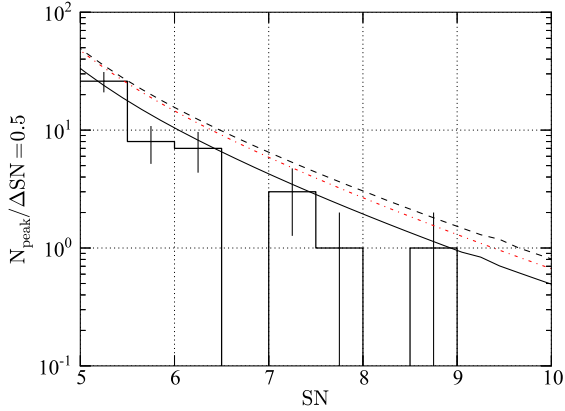}
 \end{center}
 \caption{The histogram with error bars shows the observed peak count. The solid
   line  and the dashed line present the expected peak counts from analytic
   models based on WMAP9 and Planck15 cosmology, respectively. The
   details of the analytic model is presented in Appendix~1.
   Dash-dotted line presents the peak count with Planck15 cosmology where
   dilution by member galaxies are considered (see Appendixes~2 to see how the
   dilution is modeled). 
The lines show the numbers of mass map peaks integrated over the same
bin size of $\Delta\nu=0.5$ as that used for the histogram, i.e., the
line value at each $\nu$ indicates the predicted total number of mass
map peaks between $\nu-0.5\Delta\nu$ and $\nu+0.5\Delta\nu$.
}
   \label{snhist}
\end{figure}

We compare the observed numbers of peaks with the analytic model
calculation, which is presented in Appendix~1. The analytic model is
verified by the peak numbers derived from 100 mock galaxy catalogs from 
all-sky $N$-body simulations with ray-tracing calculations
\citep{oguri17b,takahashi17}. 
The observed peak count is shown in Figure~\ref{snhist} as a histogram
whereas the peak counts expected from the analytical model assuming
the WMAP nine year cosmological parameter results 
\citep{2013ApJS..208...19H} is shown as a solid line (hereafter
WMAP9). We see a good agreement between the observation and the
model. Dashed line in the figure shows the expected peak count assuming
the Planck 2015 cosmological parameter results
\citep{2016A&A...594A..13P} whose $\sigma_8$ is higher than that of
WMAP9 (hereafter Planck15). The result suggests that our observation
favors the low $\sigma_8$ WMAP9 cosmology. 
We estimate the effect of the dilution by cluster member galaxies (see
Appendix~2 for details), and find that it is not sufficiently large to
change our conclusion. However, as discussed in Appendix~2, our
calculation may underestimate the dilution effect.  More careful
assessment of the dilution effect should be made before drawing any
firm conclusion.

\begin{figure}
 \begin{center}
  \includegraphics[width=8cm]{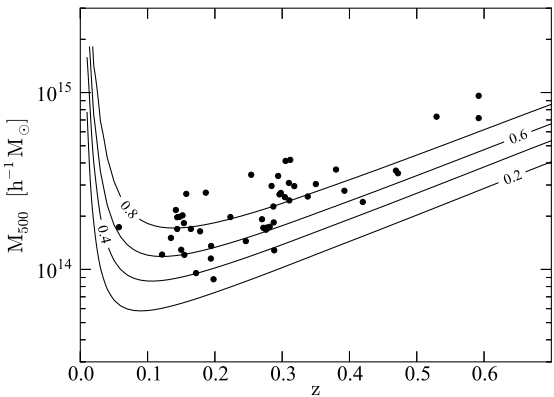}
 \end{center}
 \caption{Completeness of shear selected clusters ($\nu>4.7$) estimated from 
the analytic model, assuming the WMAP9 cosmology (Solid line contour) in the
$M_{\rm 500}$-$z$ plane (Eqn.(\ref{eqn:s})).  Filled circles show the masses and
redshifts of 
 individual peaks, where the masses are estimated from tangential shear
 fitting. Here we show only isolated peaks for which only one CAMIRA
 clusters are matched.} 
  \label{zMdiagram}
\end{figure}

\begin{figure}
 \begin{center}
  \includegraphics[width=8cm]{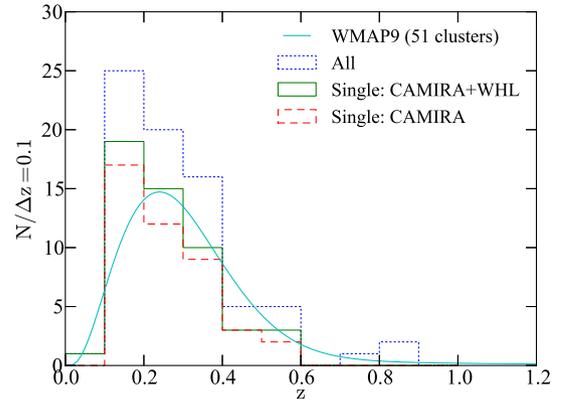}
 \end{center}
 \caption{The redshift distribution of shear selected clusters. Solid
   and dashed histograms show the observed redshift distributions for
  peaks that match CAMIRA only and CAMIRA+WHL15 catalogs,
  respectively. The dotted histogram shows the redshift distribution
  of peaks including those that have multiple counterparts in the
  CAMIRA catalog. The solid line shows the expected redshift
  distribution derived from the analytic model, with its amplitude
  normalized to the total number for the single CAMIRA case.}
  \label{nz}
\end{figure}

In addition to the comparison of the number counts, the analytic model
allows us to estimate the the selection function of the shear selected
cluster sample. Figure~\ref{zMdiagram} compares the selection function
in the the $M_{\rm 500}$-$z$ plane derived from the analytic model is
compared with estimates of $M_{\rm 500}$ for individual peaks from
fitting of tangential shear profiles presented in
Section~\ref{sec:tanshear}. We find that masses and redshifts of
individual peaks are indeed located in the relatively high
completeness region of the $M_{\rm 500}$-$z$ plane. The expected
number counts of shear selected clusters are obtained by combining the
completeness with the mass function of dark halos. Figure~\ref{nz}
compares the redshift distribution of our shear selected cluster
sample with the theoretical expectation, which again shows a good
agreement. The well-defined selection function is one of the biggest
advantages of a shear selected cluster sample.

\subsection{Weak lensing masses of X-ray clusters}
We find that two peaks have counterparts in the MCXC catalog
(Section~\ref{xraycounterpart}), both of which are Abell
clusters. Table~\ref{matchmcxc} presents the mass estimates for these
clusters, both from the tangential shear lensing signals and from the
X-ray luminosity from \citet{2011A&A...534A.109P}. Although the error associated
with X-ray mass estimate is not given in the literature, at least a level of
20\% error is expected considering the large intrinsic scatter in the
$L_X$-$M_X$ relation. Therefore, we find that both the lensing and
the X-ray mass estimates are consistent for these two clusters. 

There are two other MCXC clusters inside our survey area, for which no
counterpart is found in the shear selected cluster catalog. One is
MCXC~J0920.0+0102 at the redshift of 0.0175 whose mass estimated from X-ray
luminosity is $0.13\times 10^{14}M_\odot$. Because the redshift of
the cluster is too low, the lensing efficiency is small for this
cluster and as a result the probability to detect the clusters of
that mass is less than 10\% according to
Figure~\ref{zMdiagram}. The other one is MCXC~J1415.8+0015 at $z =
0.1259$ with the mass of $1.2\times 10^{14}M_\odot$ which 
is inconsistent with our 2$\sigma$ upper limit of 
$0.6\times 10^{14}M_\odot$. From Figure~\ref{zMdiagram}, we estimate
that the probability to detect a cluster with the mass and redshift of
 MCXC~J1415.8+0015 is 80\%. Therefore, the discrepancy between the two
 mass estimates is not negligible for this cluster. In summary, the
 weak lensing mass estimates of three out of four clusters are
 consistent with those from X-ray, whereas one cluster shows mild
 inconsistency in the mass estimates.

\begin{figure}
 \begin{center}
  \includegraphics[width=8cm]{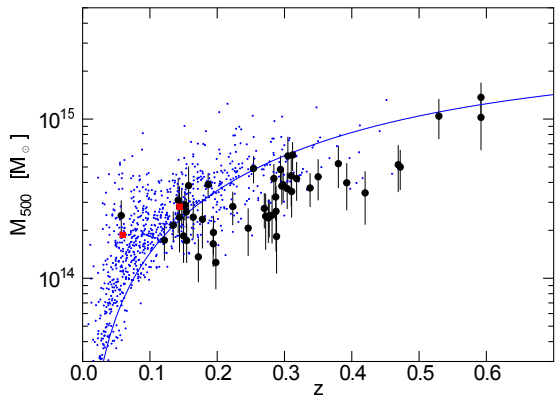}
 \end{center}
 \caption{Distributions of shear selected clusters ({\it large black
     circles}) and the NORAS/REFLEX X-ray clusters compiled in the MCXC catalog 
     ({\it small blue circles})
     on the $M_{\rm 500}$-$z$ plane . Two MCXC clusters that are
     matched with the shear selected cluster sample are shown by red
     squares. The solid line shows the typical lower mass limit of
     MCXC clusters estimated from the flux limit of $1.8\times
     10^{-12}{\rm erg\,s^{-1}cm^{-2}}$ combined with the $L$-$M$
     relation given by the equation~(10) in \citet{2014A&A...570A..31B}.}
  \label{redshiftMnfwphotoz}
\end{figure}

Figure~\ref{redshiftMnfwphotoz} presents the comparison of the shear selected
clusters with the MCXC clusters in the the $M_{\rm 500}$-$z$ plane. 
Among the shear selected clusters, those that are matched with
single CAMIRA clusters are shown here to guarantee the accuracy of the
weak lensing mass estimates. This comparison indicates that weak
lensing shear tends to probe less massive clusters toward higher
redshifts, although there is an overlapping region in which clusters
from both the cluster samples are distributed. We are particularly
interested in shear selected clusters above the typical mass limit of MCXC
estimated from the flux limit of $1.8\times 10^{-12}{\rm erg\,s^{-1}cm^{-2}}$
combined with the $L$-$M$ relation given by the equation~(10) in
\citet{2014A&A...570A..31B} (solid line in Figure~\ref{redshiftMnfwphotoz}) and
the redshift below 0.2 where mass estimates are expected be more robust. There
are five peaks in our catalog that satisfy this criterion, which are
summarized in Table~\ref{part36nomcxc}. We check the {\it XMM}, {\it Chandra}
and {\it Suzaku} X-ray archives to identify X-ray counterparts for these five
peaks, and find that peak rank 31 has its counterpart in the
Chandra archive. Because the X-ray source falls in the gap of the
Chandra CCDs, only partial image is obtained. The X-ray emission is
extended, and from the partial image we derive the lower limit of the
X-ray source to be $2.2\times 10^{43}{\rm erg\,s^{-1}}$. 

For the other four peaks (rank 3, 24, 38, 46), neither {\it Suzaku},
{\it XMM-Newton}, nor {\it Chandra} archive data are currently available. We
thus
estimate the X-ray luminosity of each cluster from the RASS spectrum
integrated from a circular region centered on the peak. The extraction
radius is $1'$ ($\sim 1$~Mpc) for rank 3, 38, 46, whereas it is $9'$
for rank 46 because of the lack of data within $6'$. Assuming that the
observed RASS spectrum is represented by a sum of ICM thermal
emission, the Galactic soft X-ray emission (0.1~keV Local Hot Bubble
and 0.3~keV Milky Way Halo components), and the comic X-ray background
\citep{kushino02}, we perform the spectral fitting and derive the
$3\sigma$ upper limit on the 0.1--2.4~keV X-ray luminosity
(Table~\ref{part36nomcxc}). Here the APEC thermal plasma model
\citep{smith01} is assumed for the ICM component and the temperature
and the metal abundance are fixed at 5~keV and 0.3~solar, respectively. 
From a comparison with the mass-luminosity relations given in Figure~3
of \citet{2015MNRAS.447.3044G}, it appears that shear-selected
clusters tend to be underluminous in X-ray. Due to large uncertainties
associated with the present analysis, however, 
further deep observations using {\it Chandra} or {\it XMM-Newton} would be
essential to constrain X-ray temperature and other properties. 
In fact, inside the
search radius of $3'$, two peaks, rank 24 and 38, are found to have
faint X-ray sources on the second ROSAT all-sky survey catalog
\citep{boller16}; 2RXSJ162010.5+422811, 2RXSJ120955.7-003234,
respectively.  

\begin{table}[h]
\caption{Massive shear selected clusters that has no counterpart in the MCXC
X-ray cluster catalog}
\vskip 2mm
\begin{tabular}{llll} \hline
rank & $z$ & $M_{\rm 500, WL}$ & $L_{500, X}$ \\
      &           & [$\times 10^{14}$M$_\odot$] & [$\times 10^{43}$erg/s] \\ \hline
3     & 0.1540 &  2.6$\pm$0.6   & $<$ 6 \\
24    & 0.1473 &  3.1$\pm$1.2   & $>$ 2.2\\
31    & 0.1420 &  2.8$\pm$0.8   & $<$ 6 \\
38    & 0.1780 &  2.3$\pm$0.8   & $<$ 8 \\
46    & 0.1575 &  3.9$\pm$1.2   & $<$ 8 \\
\hline
\end{tabular}
\label{part36nomcxc}
\vskip 1mm
\begin{tabnote}
We find the counterpart of the peak 32 on Chandra archive. All the other upper limit
come from RASS. 
\end{tabnote}
\end{table}

We then compare the masses of the four clusters that have matched with
the published X-ray clusters in the XXL survey. The mass estimate XXL
cluster is based primarily on the X-ray temperature inside the fixed
300~kpc aperture which was calibrated by the weak lensing data taken
by CFHT Legacy Survey \citep{xxl4}. The results shown
in Table~\ref{matchxxl} indicate that our weak lensing mass estimates
are consistent with the XXL mass estimates for the four clusters.
This is somehow inconsistent with the suggestion by
\citet{2015MNRAS.447.3044G} that weak lensing selected clusters tend
to be embedded in the filaments viewed near the line-of-sight and the
mass tends to be overestimated by the filaments. 
\citet{2012MNRAS.425.2287H} also argued, based on ray-tracing
simulations and an analytic halo model, that shear selected clusters
exhibit the orientation bias such that their major axes are
preferentially aligned with line-of-sight direction, which also
induces the mismatch between lensing and X-ray masses. We see no
evidence of such a bias in our weak lensing selected samples shown
above, presumably due to the small sample size.

\begin{figure}
 \begin{center}
  \includegraphics[width=8cm]{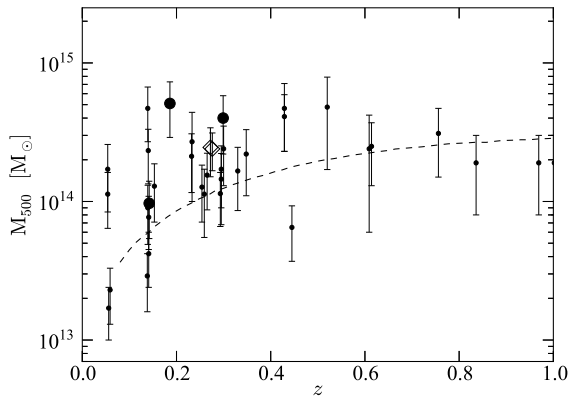}
 \end{center}
 \caption{The distribution of XXL clusters in the $M_{\rm 500}$-$z$ plane.
   Small filled circles indicate XXL clusters inside our HSC S16A
   survey area, whereas large filled circles are XXL clusters that are
   matched with the lensing peaks. Open diamond show the redshifts and
   lensing masses of shear selected clusters that do not have
   counterparts in the XXL cluster catalog. Dashed line shows the 50 \%
  completeness limit of the XXL cluster sample.}  
  \label{xxlpart36}
\end{figure}

In the cross-match of the peaks with the published XXL clusters, there
are two peaks that do not have counterparts in the XXL cluster
list. We show these samples in the $M_{\rm 500}$-$z$ plane in
Figure~\ref{xxlpart36}, together with other XXL clusters inside our
HSC S16A survey area. Both  peaks are located above the 50\%
completeness limit of the XXL cluster sample. Therefore, given their
lensing masses these clusters should have been detected by XXL,
assuming that these clusters follow the standard mass-luminosity
scaling relation. We note, however, that the XXL cluster catalog was constructed 
by using both X-ray core size and flux, and therefore such non-detected clusters
might be more extended rather than under-luminous. 
Further studies over wide field is vitally important to understand X-ray gas
properties and also to study some other systematic bias in different techniques
to search clusters. 
The {\it eROSITA} will provide a good opportunity to conduct this type of work.  

\begin{figure*}
 \begin{center}
  \includegraphics[width=8cm]{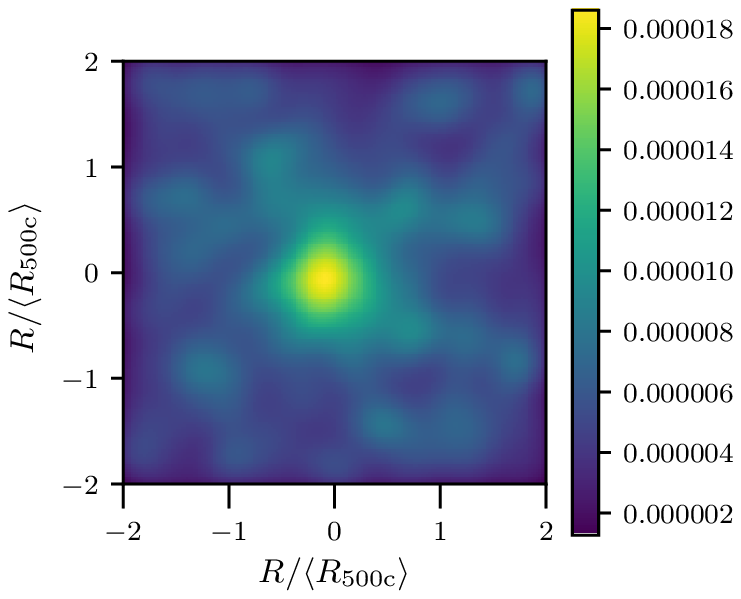}
  \includegraphics[width=8cm]{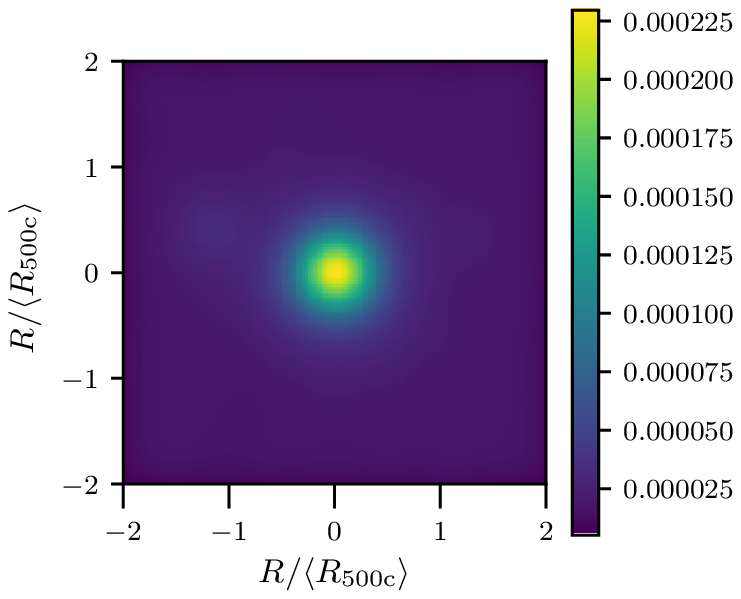}
 \end{center}
	\caption{ The stacked $0.5-2.0$ keV RASS images of shear selected clusters (left panel) and MCXC selected clusters (right panel) selected to have X-ray luminosity greater than that expected for our mass threshold. The images have been smoothed by a 9 pixel gaussian kernel. The scale in the left (right) panel correspond to the mean counts per second expected from a cluster that went into the stack at the median redshift of $z=0.27$ ($z=0.14$).
  }  
  \label{rosat_stack}
\end{figure*}

\subsection{Stacked X-ray emission}\label{sec:stackedX}

As we were able to identify only a limited number of X-ray counterparts for the
shear-selected peaks, we stack the X-ray emission from {\it ROSAT} around the shear
selected peaks with a procedure somewhat similar to \citet{Anderson:2015}. As
the shear peaks have limited centering accuracy, we used the position and
redshift of the optical cluster counterparts that correspond to our
shear-selected peaks. We first extract a RASS image in the energy band
(0.5-2.0) keV around the optical counterparts extending out to the mean
$2R_{\rm 500c}$, as well as the corresponding RASS exposure map. Dividing the
RASS image by the exposure map, yields an image with the counts per second for
each cluster. We then perform a weighted addition of these RASS images together
to produce a stacked image. The weight we use $D_{\rm L}^2(z)/D_{\rm
L}^2(z_{\rm med})$, standardizes the flux to its expectation at the median
redshift of our shear selected clusters ($z_{\rm med}=0.27$).  Known point
sources are excluded in the stacking procedure. Such a stacked image is shown
in Figure~\ref{rosat_stack} and shows that we can clearly detect emission from
our shear-selected clusters once they are stacked. We have checked that our
stacked image is similar even if we first stack the raw count images and divide
by the total exposure at the end.

We then compute the count rate of X-ray photons within $R_{\rm 500c}$
for the average mass and subtract the expected count rate due to the background estimated within an
annulus of $[1, 2]R_{\rm 500c}$. We obtain a net count rate $(3.2\pm0.6)\times 10^{-2}$ cnt/s,
where the errors on the count rates were estimated using the jackknife
technique. These count rates are then converted to a X-ray luminosity assuming
a metallicity which is 0.2 solar, the average column density $\langle N_H
\rangle=3\times10^{20}{\rm cm}^{-2}$.
The X-ray luminosity we obtain for our shear selected clusters is $L_{\rm
x}=(1.5\pm0.3)\times 10^{44} {\rm ergs^{-1}}$ in the $0.1-2.4$keV.

To compare with X-ray selected clusters, we similarly stacked the RASS images
around clusters in MCXC catalog \citet{2011A&A...534A.109P} with an X-ray
luminosity greater than that expected given the mass-detection threshold for
our shear-selected clusters (see Figure \ref{zMdiagram}).  We also applied the
following selection criteria $0.01<N_H/10^{22}{\rm cm^{-2}}<0.06$ and
$0.01<z<0.6$ which corresponds to the same range as those for shear-selected
clusters.  The average X-ray luminosity of these clusters based on the MCXC
catalog is $L_{\rm x}=2.9 \times 10^{44} {\rm ergs^{-1}}$ in the $0.1-2.4$keV
band.  Our procedure for computing $L_{\rm x}$ based on the stacked RASS image
(see right panel of Figure~\ref{rosat_stack}) yields $(29\pm2)\times 10^{-2}$
cnt/s at a median redshift of $z=0.14$. This corresponds to $L_{\rm
x}=(3.1\pm0.2)\times 10^{44} {\rm ergs^{-1}}$ in agreement with the average
value based on the MCXC catalog.  This shows that the X-ray luminosity of our
shear-selected clusters is about half of that expected from X-ray selected
clusters at $\sim5\sigma$ level.  Since the shear-selected clusters are
unbiased to selection effects in X-ray surveys concerning X-ray luminosity and
dynamical states, this demonstrates its potential power to discover X-ray
underluminous clusters. Once selection effects due to orientation biases are
accounted, the shear selected clusters could potentially provide hints to
resolving the discrepancy in the number of clusters discovered through the SZE
\citep{2016A&A...594A..24P} or X-ray \citep{xxl2} method and the predictions
for these abundances based on cosmological parameters inferred from the cosmic
microwave background observations \citep{2016A&A...594A..13P}.

\subsection{Average radial density profile of clusters from stacked lensing} \label{avradialprofile}

The lensing signal depends both on the total mass of the cluster and its radial
density profile. More concentrated clusters are more likely to be detected by
lensing. In order to see how our weak lensing clusters sample differs from
clusters studied so far, we check the average radial density profile
of our sample. As is presented in Figure~\ref{rdsigmarank4}, lensing
signals from individual clusters are not sufficient to constrain the
radial profile accurately. Instead, we carry out a stacked lensing
analysis. We compute the differential surface mass density
(equation~\ref{deltasigmaR}) for each cluster. We select source
galaxies behind the cluster based on the photometric redshift. 
We then stack them to obtain the average differential surface mass density
\begin{equation}
  \Delta \Sigma(R) = \left< \Sigma_{\rm cr,i} \gamma_{T, i}(R = D_{l,i}\theta)\right>
\end{equation}
where we calculate averaged $\Sigma_{\rm cr}$ for each $i$-th cluster and the
angular separation from the center, $\theta$, is scaled to physical length when
stacking. 

\begin{figure}
 \begin{center}
  \includegraphics[width=8cm]{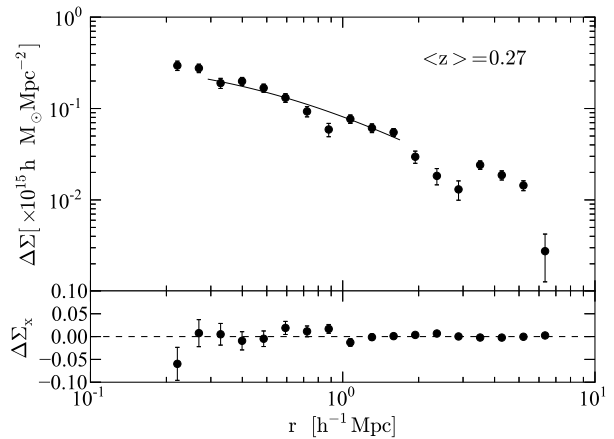}
 \end{center}
 \caption{Stacked differential surface density profile from about 50
   shear selected clusters. The solid line shows the best-fit model
   assuming the NFW density profile. Fitting is made in the physical
   scale between $0.3$ to $1.7h^{-1}$Mpc. Photometric redshifts of
   individual galaxies are used to select background galaxies. The
   best-fit values are ($M_{500}$, $c_{500}$) = 
   (($2.03\pm0.13)\times 10^{14}h^{-1}M_{\odot}$, $1.91\pm0.37$).
Bottom panel shows the profile generated by B-mode components.}
\label{rdsigmaall}
\end{figure}

\begin{figure}
 \begin{center}
  \includegraphics[width=8cm]{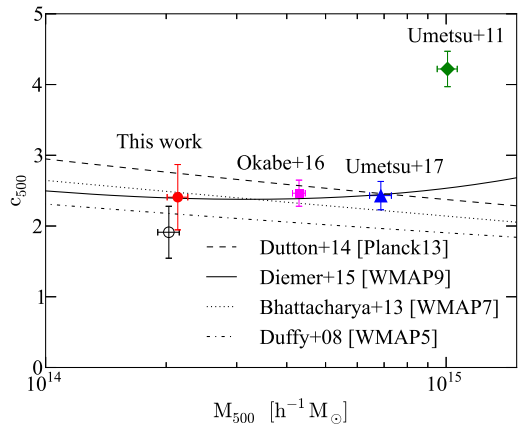}
 \end{center}
 \caption{Constraints on the mass-concentration relation. The result of
   the fitting of the NFW profile to the stacked differential surface
   density for about 50 shear selected clusters (see
   Figure~\ref{rdsigmaall}) is shown by the open circle. After the dilution
   correction (see text), the constraint is shifted to the point shown by the
   filled circle.  The average redshift of the shear selected clusters
   is 0.27. 
The filled triangle shows the results for 16 X-ray-selected clusters at an
average redshift of 0.34 obtained from a strong and weak lensing
analysis of \citet{umetsu17}. The filled diamond shows the results for
a sample of 4 strong-lensing-selected superlens clusters at an average
redshift of 0.32 from a strong and weak lensing analysis of \citet{umetsu11}.
The filled square is from \citet{okabe16} estimated from 50 X-ray luminous
(LoCuSS) clusters at redshift between 0.15 and 0.3
}
\label{mcrelationstack}
\end{figure}

Figure~\ref{rdsigmaall} shows the stacked differential surface density profile
around 50 weak-lensing peaks that are matched with single isolated CAMIRA
clusters. The average redshift of the sample is $0.27$. The data is
fitted to an NFW profile on the physical scale between $0.3$ to
$1.7h^{-1}$Mpc to obtain the mass, $M_{500}$ and the concentration
parameter, $c_{500}$ as $(2.03\pm0.13)\times 10^{14}h^{-1}M_{\odot}$
and $1.91\pm0.37$, respectively. Here we adopted $1.7h^{-1}$Mpc for the outer
radius limit to avoid possible contamination from 2-halo term. 
This result is shown in Figure~\ref{mcrelationstack} as a open circle.

In projection, the 2-halo term is nearly constant out to about 3 virial radii,
so that the 2-halo contribution to the differential surface mass density
($\Delta\Sigma$) is negligible at small cluster-centric distances, and it becomes
dominant only at large distances, R $\sim$ 5 virial radii \citep{ogurihamana11}.
These expected behaviors are
consistent with recent cluster weak-lensing observations (e.g. \cite{okabe16}),
and the NFW density profile provides an excellent
description of the projected matter distribution around clusters out
to about 2 virial radii \citep{umetsu16}. Therefore, the 2-halo term with in
the physical scale of $1.7h^{-1}$ Mpc is negligible.

If cluster member galaxies are mis-identified as background galaxies,
the observed lensing signal can be systematically underestimated
progressively toward the cluster center \citep{medezinski10}, 
in proportion to the fraction of unlensed galaxies whose
orientations are randomly distributed. This dilution effect is a major
source of cluster weak-lensing measurements, and is not accounted for
in our analysis described above. To establish optimal source selection
schemes for unbiased cluster weak-lensing measurements,
\citet{medezinski17} examined the degree of cluster contamination by
analyzing a large sample of $\sim 900$ CAMIRA clusters with richness
greater than 20 identified in $\sim 160$\,deg$^2$ of HSC-SSP data.
According to the findings of \citet{medezinski17} (see Figure 10 of
\cite{medezinski17}), the level of dilution reaches up to $15\%$
at a cluster-centric radius of $r=0.3 h^{-1}$\,Mpc and becomes less
than $10\%$ at $r \gtrsim 0.5h^{-1}$\,Mpc.\footnote{We note that
\citet{medezinski17} measured the weak-lensing signals around CAMIRA
clusters as a function of comoving cluster radius, whereas we stack
clusters in proper length units.} 
We account for this dilution effect
by applying corrections obtained by \citet{medezinski17} at each
cluster radius $r$. The filled circle in Figure~\ref{mcrelationstack}
shows the best-fit $c$--$M$ relation ($M_{500}=(2.13\pm0.13)\times
10^{14}h^{-1}M_{\odot}$, $c_{500}=2.41\pm0.46$) derived from our
stacked lensing analysis after correcting for the dilution effect.
Compared with the original estimates (open circle), the dilution correction
increases $M$ and $c$ by 5 \% and 26 \%, respectively. Note that $M$ and $c$ in
Table~\ref{papertablephotoz} are the values before the dilution correction.

Mis-centering of dark matter concentration can also lower estimates of the
concentration parameter. Here, we adopted the peak position as the
center, whose uncertainty is expected to be an order of the smoothing
scale that we adopted, i.e., $\sim 1'$. This angular scale is smaller
than the innermost radius of radial profile fitting, $0.3h^{-1}$~Mpc,
for the  clusters below the redshift of 0.5 where most of our samples are
located. Thus the mis-centering does not affect the estimate of the
concentration parameter significantly. 

It is interesting to adopt the CAMIRA's BCG positions as the center of halo and
to try the same stacking analysis. The best fit parameter that obtained through
this trial is ($M_{500}$, $c_{500}$) = 
($1.78\pm0.12)\times 10^{14}h^{-1}M_{\odot}$, $1.13\pm0.23$).
Compared with the original estimates, the mass does not change significantly
whereas $c$ drops largely. This suggests that the adoption of the BCG position
in fact increase the degree of mis-centering of the dark matter halos,
which is consistent with the expectation as discussed in Sec.~\ref{opticalcounterparts}.

Several groups have reported observational results from NFW fits to
the lensing profiles of X-ray luminous clusters \citep{umetsu09, okabe16, umetsu17}.
They obtained $c_\mathrm{vir} \sim 6$ for 
$M_\mathrm{vir}\sim 10^{15}h^{-1}M_\odot$ clusters, which corresponds to
$c_\mathrm{500}\sim 3$ and is consistent with our results. 
\citet{2006MNRAS.372..758M} estimated the mass-concentration relation
using galaxy-galaxy lensing based on the SDSS data and showed that
$c_{\rm vir} = 5\pm 1 $ ($c_{500}\sim 2.5$) in our mass
range. Therefore, our estimate is also consistent with their result.  

On the other hand, \citet{umetsu11} obtained $c_\mathrm{vir} =
7.68^{+0.42}_{-0.40}$ from spherical NFW fits for a sample of four
strong-lensing-selected {\em superlens} clusters with 
$M_\mathrm{vir}\simeq 1.5\times 10^{15}h^{-1}M_\odot$, corresponding
to $c_\mathrm {500}=4.22\pm 0.25$ at $M_\mathrm{500}=(10.08\pm
0.58)\times 10^{14}h^{-1}M_\odot$. The superlens sample is
characterized by large Einstein radii $> 30$ arcsec (for a fiducial
source redshift of $z=2$). Their concentration is apparently higher
than our $c$ measurement of the shear selected 
clusters. \citet{2012MNRAS.420.3213O} found similarly high
concentration parameter estimates for their strong-lensing clusters
selected from SDSS data. This discrepancy can be solely explained by
the selection and projection bias of strong-lensing-selected clusters
as discussed in \citet{2012MNRAS.420.3213O} and \citet{umetsu17}.

Since the lensing-derived concentration parameter is sensitive to the
selection and projection effects, if we obtain a halo concentration
that is consistent with the theoretical mass-concentration relation,
it may imply that shear selected cluster populations of clusters are
statistically different from those selected by the strong-lensing
properties and they are not biased toward very high concentrations.
Unlike strong-lensing-selected cluster samples, our weak lensing
clusters are selected primarily because they are massive.
While \citet{2012MNRAS.425.2287H} predicted the moderate orientation
bias for shear selected clusters, the bias is predicted to be much
weaker than that for strong-lensing-selected clusters. Therefore we may
need a much larger sample of shear selected clusters to find any
possible evidence of the selection bias for shear selected clusters.
A caveat is that \citet{2012MNRAS.425.2287H} predicted larger
orientation bias for higher $\nu$ peaks, which should also be tested
with observations.

In Figure~\ref{mcrelationstack}, several numerical predictions are
shown, including \citet{duffy08}, \citet{dutton14},
\citet{bhattacharya13}, and \citet{diemer15}, where different models
are calibrated for different $\Lambda$CDM cosmologies. For the
\citet{diemer15} model, which is applicable for any cosmology, we
assume the WMAP9 cosmology. Combined with the result with
\citet{umetsu17} and taking the dilution correction into account, we
slightly disfavor the prediction by \citet{duffy08} which assumes
WMAP5 in their calculation.  

The cluster mass-concentration relation is sensitive to cosmological
parameters. In particular, its normalization is highly sensitive to
the combination of $\Omega_\mathrm{m}$ and $\sigma_8$, As demonstrated
in \citet{dutton14}, more recent cosmological models (since WMAP3)
take higher values of ($\Omega_\mathrm{m}, \sigma_8$), thus predicting
a higher normalization in the mass-concentration relation. This may
explain the inconsistency against \citet{duffy08} adopting WMAP5 whose
normalization is lower than the later WMAP and Planck.  

Among the concentration models considered here, all except
\citet{diemer15} show a monotonic decrease of $c$ as $M$ increases.  
\citet{diemer15}, however, predicts an upturn at high masses due to
dynamically un-relaxed, forming halos \citep{ludlow10}. 
With the full survey data in hand where we expect more than a six times larger
sample, we will be able to distinguish these concentration models and to
observationally establish the mass-concentration relation over a wide
range of the cluster mass and the redshift. 

\section{Conclusion}\label{sec:conclusion}
We have constructed a shear selected cluster catalog containing 65
mass map peaks with $\nu>4.7$ from the HSC-SSP Wide S16A dataset
covering $\sim 160$~deg$^2$. The cluster catalog is large in size yet
has small contaminations from false positives. We have found that all
the mass map peaks with $\nu > 5.1$ have physical counterparts, and
only 2 out of the 65 clusters are probably false positives. 
The peak counts are found to be consistent with the predictions from
WMAP9 cosmology but the tension appears to exist if we adopt the
Planck15 cosmology. The comparison with the published X-ray catalogs
have shown that cluster masses estimated from lensing are consistent
with X-ray mass estimates, although the matched sample is small. On
the other hand, there exist a few X-ray undetected shear selected
cluster candidates which should have been detected by the existing X-ray
surveys if the clusters follow the standard mass-luminosity relation.
We stacked RASS X-ray images around the shear-selected clusters and
compare with X-ray selected clusters given the same selection function
of the mass. We found that the average X-ray luminosity for the shear-selected
clusters are about half of that of the X-ray selected clusters at the
$\sim5\sigma$ level. It indicates the existence of the population of
X-ray underluminous clusters and that the shear-selection can discover such a
population. 
Future studies to constrain the abundance of such clusters
are necessary. Stacked lensing analysis suggests that the average
radial  mass profile of the shear selected clusters is not too
concentrated, which implies that shear selected clusters do not
exhibit strong selection bias on the orientation or the internal
structure, and that the clusters are selected basically because they
are massive. 

This paper presents the initial results on shear selected clusters from the
HSC-SSP survey. At the completion of the HSC-SSP survey, the shear selected
cluster sample would potentially be at least six times larger, which is
sufficiently large to conduct various statistical studies with high accuracy.

\begin{ack}
This work was supported in part by World Premier International Research Center Initiative (WPI Initiative), MEXT, Japan, and JSPS KAKENHI Grant Number 15H05892, 26800093.
MO and SM acknowledges financial support from JST CREST Grant Number JPMJCR1414. SM is also supported by the Japan Society for Promotion of Science grants JP15K17600 and JP16H01089.

The Hyper Suprime-Cam (HSC) collaboration includes the astronomical communities of Japan and Taiwan, and Princeton University.  The HSC instrumentation and software were developed by the National Astronomical Observatory of Japan (NAOJ), the Kavli Institute for the Physics and Mathematics of the Universe (Kavli IPMU), the University of Tokyo, the High Energy Accelerator Research Organization (KEK), the Academia Sinica Institute for Astronomy and Astrophysics in Taiwan (ASIAA), and Princeton University.  Funding was contributed by the FIRST program from Japanese Cabinet Office, the Ministry of Education, Culture, Sports, Science and Technology (MEXT), the Japan Society for the Promotion of Science (JSPS),  Japan Science and Technology Agency  (JST),  the Toray Science  Foundation, NAOJ, Kavli IPMU, KEK, ASIAA,  and Princeton University.

The Pan-STARRS1 Surveys (PS1) have been made possible through contributions of the Institute for Astronomy, the University of Hawaii, the Pan-STARRS Project Office, the Max-Planck Society and its participating institutes, the Max Planck Institute for Astronomy, Heidelberg and the Max Planck Institute for Extraterrestrial Physics, Garching, The Johns Hopkins University, Durham University, the University of Edinburgh, Queen's University Belfast, the Harvard-Smithsonian Center for Astrophysics, the Las Cumbres Observatory Global Telescope Network Incorporated, the National Central University of Taiwan, the Space Telescope Science Institute, the National Aeronautics and Space Administration under Grant No. NNX08AR22G issued through the Planetary Science Division of the NASA Science Mission Directorate, the National Science Foundation under Grant No. AST-1238877, the University of Maryland, and Eotvos Lorand University (ELTE).
 
This paper makes use of software developed for the Large Synoptic Survey Telescope. We thank the LSST Project for making their code available as free software at http://dm.lsst.org.

Based in part on data collected at the Subaru Telescope and retrieved from the
HSC data archive system, which is operated by the Subaru Telescope and Astronomy
Data Center at National Astronomical Observatory of Japan.

HM is supported by the Jet Propulsion Laboratory, California Institute of
Technology, under a contract with the National Aeronautics and Space
Administration.
\end{ack}

\section*{Appendix 1. Expected peak counts from weak lensing SN-map}

Here we present a halo model prediction of the number counts of peaks,
focusing on the high $\nu$ region where single clusters dominate the
signals. 

The signal to noise ratio $\nu$ defined in equation~(\ref{defnu})
considers only the pure statistical noise of the measurement,
$\sigma_{\rm noise}$, which is proportional to
$\sigma_e/\sqrt{N_{gal}}$, where $\sigma_e$ and $N_{gal}$ is the
intrinsic galaxy ellipticity per component and the number of source
galaxies, respectively.  In the real cluster search with weak lensing
mass peaks, the large-scale structure along the line-of-sight acts as
a noise, which we describe as $\sigma_{\rm LSS}$. It can be computed
from the cosmic shear power spectrum $C_\ell$ as
\begin{equation}
\sigma_{\rm LSS}^2=\int \frac{\ell
  d\ell}{2\pi}|\hat{U}(\ell)|^2C_\ell,
\label{sigmalss}
\end{equation}
where $\hat{U}(\ell)$ is the Fourier counterpart of the filter
$U(\theta)$ 
\begin{equation}
\hat{U}(\ell)=2\pi\int\theta d\theta U(\theta)J_0(\ell\theta). 
\end{equation}
The cosmic shear power spectrum is related to the matter power
spectrum $P_m(k;z)$ as 
\begin{equation}
C_\ell =\int d\chi
\frac{\left[W^\kappa(\chi)\right]^2}{\chi^2}P_m(k=\ell/\chi; z),
\end{equation}
\begin{equation}
W^\kappa(\chi)=\frac{\bar{\rho}(z)}{(1+z)\Sigma_{\rm cr}(z)},
\end{equation}
where the critical surface density $\Sigma_{\rm cr}$ is
defined in equation~(\ref{sigmacrit}). 
We fully take account of the redshift distribution of source galaxies
by computing $\langle D_{ls}/D_s\rangle$ by averaging the distance 
ratios over the sum of the probability distribution functions of 
photometric redshifts of individual galaxies \citep{tanaka17}.
For the matter power spectrum $P_m(k;z)$, we use the halofit model 
of \citet{smith03} and \citet{takahashi12}. 

For a single NFW halo with the mass $M$, the concentration parameter
$c$, and redshift $z$, the peak SN $\nu_{\rm peak}$ is computed as
\begin{equation}
  \nu_{\rm peak} = \frac{\kappa_{\rm peak}(M,c,z)}{\sigma_{\rm noise}},
  \label{nupeak}
\end{equation}
\begin{equation}
\kappa_{\rm peak}(M,c,z)=2\pi \int d\theta \theta \kappa_{\rm NFW}(\theta; M,c,z)U(\theta),
\end{equation}
where $\kappa_{\rm NFW}$ is the convergence profile for the NFW
halo. Again, we include the redshift distribution of source galaxies
in $\Sigma_{\rm cr}$ that is used to compute $\kappa_{\rm NFW}$.
We then compute the number density of mass map peaks with
$\nu_1<\nu<\nu_2$ as 
\begin{eqnarray}
  \frac{d N}{d\Omega}(\nu_1,\nu_2)&=&\int dz \frac{dV}{dzd\Omega}\int dM \frac{dn}{dM}
  \int dc \frac{dp}{dc} \nonumber\\
  &&\times S(M,z,c;\nu_1,\nu_2),
\end{eqnarray}
where $dn/dM$ is a mass function of dark halos \citep{tinker08},
$dp/dc$ is the distribution of the concentration parameter which we
assume log-normally distributed
\begin{equation}
 \frac{dp}{dc}=\frac{1}{\sqrt{2\pi}\sigma_{\ln
     c}}\exp\left[-\frac{(\ln c - \ln \bar{c})^2}{2\sigma_{\ln
       c}^2}\right]\frac{1}{c}.
\label{eqn:dpdc}
\end{equation}
We use the model of the concentration parameter presented by
\citet{diemer15}, which takes account of the cosmology dependence of
the concentration parameter as a function of halo mass and redshift.
The selection function $S(M,z,c;\nu_1,\nu_2)$ is
\begin{equation}
S(M,z,c;\nu_1,\nu_2)=\frac{1}{2}\left[{\rm
    erfc}\left(x_1\right)-{\rm erfc}\left(x_2\right)\right],
\label{eqn:s}
\end{equation}
\begin{equation}
x_i=\frac{1}{\sqrt{2(1+\sigma_{\rm LSS}^2/\sigma_{\rm
      noise}^2)}}(\nu_i-\nu_{\rm peak}), 
\end{equation}
where we consider $\sigma_{\rm LSS}$ (equation~\ref{sigmalss}) to
incorporate the noise enhancement by large-scale structure. 
The selection function $S(M, z, c; \nu_1, \nu_2)$ take accounts of the
scatter of peak heights due to the shot noise ($\sigma_{\rm noise}$)
as well as random projections of halos along the line-of-sight
($\sigma_{\rm LSS}$). Given that this model includes the situation that
$\nu$ of halos with relatively small masses are significantly boosted by
$\sigma_{\rm LSS}$, peaks generated by projections of multiple small
halos along the line-of-sight are taken into account in this model,
at least to some extent.

In a simple case where the source galaxy density is uniform across the
survey area, the shot noise $\sigma_{\rm noise}$ in equation~(\ref{nupeak})
can be simply estimated as
\begin{equation}
\sigma_{\rm noise}=\sigma_e\sqrt{\frac{A}{n_{\rm g}}},
\end{equation}
\begin{equation}
A=2\pi\int d\theta \theta Q^2(\theta),
\end{equation}
where $\sigma_e$ is the rms error on the galaxy ellipticity per
component. In practice, however, $\sigma_{\rm noise}$ is not uniform
but has a spatial pattern due to the effects of the boundary, masking,
and the inhomogeneous source number density. The spatial pattern of
the noise $\sigma_{\rm sigma}$ can be estimated from the variance in
mass maps using randomized galaxy shape catalogs, as done in
Section~\ref{sec:mapmake}. This allows us to derive the mass map area
as a function of $\sigma_{\rm noise}$. Specifically, we denote
$\Omega_i$ as the total area of mass maps that have the estimated
shot noise
$\sigma_i-\Delta\sigma/2<\sigma(\boldsymbol{\theta}_0)<\sigma_i+\Delta\sigma/2$.
In this case, the total number of mass map peaks with
$\nu_1<\nu<\nu_2$ is given by
\begin{equation}
N_{\rm peak}(\nu_1,\nu_2)=\sum_i \left.\frac{dN}{d\Omega}(\nu_1,\nu_2)\right|_{\sigma_{\rm noise}=\sigma_i}\Omega_i.
\label{model_ntot}
\end{equation}

\begin{figure}
 \begin{center}
  \includegraphics[width=8cm]{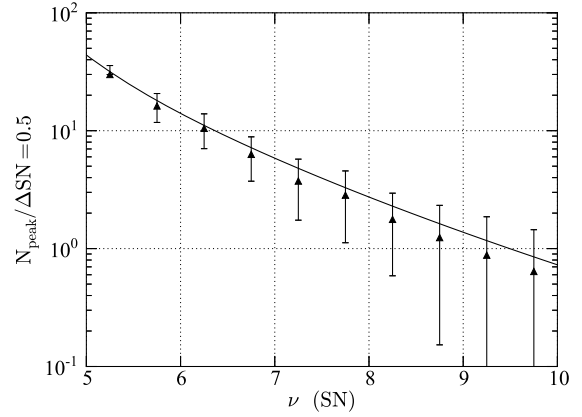}
 \end{center}
 \caption{Model predictions of the number counts of mass map peaks as
   a function of $\nu$. The solid line shows the analytic calculation
   (equation~\ref{model_ntot}), whereas filled triangles with errors
   show the peak counts from mass maps with 100 mock shear catalogs
   \citep{oguri17b}. In both analytic and mock calculations, WMAP9
   cosmology is assumed, and the multiplicative bias is set to $m=0$. }
   \label{snhistmodel}
\end{figure}

To check the accuracy of this analytic model prediction, we compare
the number counts from equation~(\ref{model_ntot}) with results from
mock shear catalogs. \citet{oguri17b} constructed mock shear catalogs
for the first-year HSC shear catalog. The mock has the same spatial
distribution as the real shear catalog from the observations, but the
ellipticities of individual galaxies with mock ellipticity values.
The mock ellipticity values are determined first by randomly rotating
the orientations of individual galaxies and then adding cosmic shear
taken from ray-tracing simulations \citep{takahashi17}. The comparison
is given in Figure~\ref{snhistmodel}. In this comparison, we assume
WMAP9 cosmology for both the analytic and mock calculations. The
multiplicative bias is also set to $m=0$ for this comparison. We find
a good agreement between the analytic model and mock result. 

\begin{figure}
 \begin{center}
  \includegraphics[width=8cm]{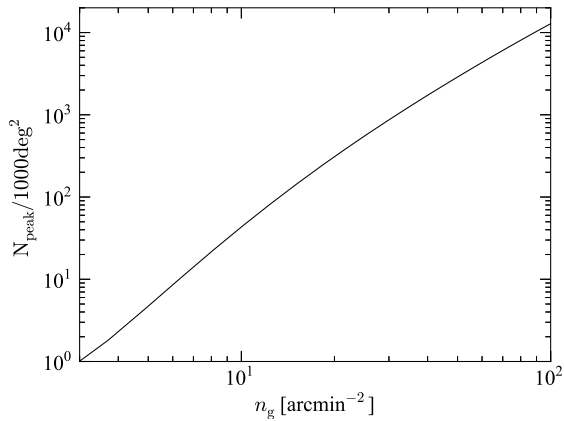}
 \end{center}
 \caption{The expected number of significant ($\nu > 5$) mass map peak
   per 1000~deg$^2$ as a function of the source number density.
   In this calculation, we adopt a uniform $\sigma_{\rm noise}$, which
   is computed from the input source number density $n_{\rm g}$ and the
   intrinsic ellipticity of $\sigma_e=0.4$. In the calculation we also
   include the evolution of the mean source redshift as a function of
   $n_{\rm g}$ assuming the relation $n_{\rm g}=30\bar{z}_m^3$arcmin$^{-2}$.}
   \label{pc_ngal}
\end{figure}

The analytic model allows us to explore the expected number density of
mass map peaks as a function of various survey parameters. As a
specific example, we predict the number density of mass map peaks with
significant SN, $\nu>5$, as a function of the survey depth. Deep imaging
increase both the source number density $n_{\rm g}$ and the mean
source redshift $\bar{z}_m$. We include this correlation assuming a
simple monotonic relation between these two as
$n_{\rm g}=30\bar{z}_m^3$arcmin$^{-2}$, which roughly reproduces the
observed trend. We also include the redshift distribution of the
source galaxies assuming $n(z)\propto z^2\exp(-z/z_0)$ with
$z_0=\bar{z}_m/3$. The result shown in Figure~\ref{pc_ngal} indicates
that the peak number counts is indeed a very steep function of the
survey depth. For example, the peak number density is enhanced by a
factor of $\sim 20$ from $n_{\rm g}=10$~arcmin$^{-2}$ to
$n_{\rm g}=30$~arcmin$^{-2}$. This prediction highlights the
importance of deep imaging surveys as realized by the HSC-SSP survey
for constructing a large sample of shear selected clusters.

\section*{Appendix 2. Dilution effect by cluster member galaxies}

Member galaxies in clusters of galaxies do not contribute to lensing
signals, and therefore dilute weak lensing signals. This is
potentially one of the most significant sources of systematic effects
in cluster weak lensing studies, and therefore has to be studied
carefully. For instance, \citet{2016MNRAS.463.3653K} studied the
dilution effect and found relatively large corrections of $\nu$ of
$\sim 10$\% around $\nu\sim 4$ due to the dilution effect, for the
case of Dark Energy Survey which is much shallower than the HSC-SSP.  
Here we employ a simple model of the number counts of cluster member
galaxies to check the impact of the dilution effect on our results. 

We use a model of the number of satellite galaxies as a function of
the halo mass and redshift derived in \citet{lin04} and \citet{lin06}
to estimate the dilution effect. We assume that the number density
profile follow the NFW profile. Since these galaxies do not contribute
to the lensing signals, the enhancement of the number density with
respect to the average density represents the enhancement of the noise
$\sigma_{\rm noise}$. We compute the enhancement of $\sigma_{\rm
  noise}$ as a function of the halo mass and redshift by convolving
the number density profile of satellite galaxies with the filter
function $Q(\theta)$, and include this in the halo model calculation
described in Appendix~1. We find that the dilution effect is modest,
with $\sim 4\%$ decreases of $N_{\rm peak}$ at $\nu=5$, and $\sim
10\%$ at $\nu=7$. Figure~\ref{snhist} indicates that the dilution
effect is not large enough to explain the apparent difference between
the observed number counts and the halo model prediction based on the
Planck15 cosmology. 

However, this is a preliminary result based on the simple model of the
number distribution of satellite galaxies, which may have room for
improvement. We also ignored the intrinsic alignment of cluster member
galaxies. Cluster member galaxies tend to be radially aligned with
respect to the cluster center, which effectively produces negative
peaks in weak lensing analysis. Thus, the inclusion of the radial
alignment may further increase the impact of the cluster member
galaxies. 

One possible way to mitigate the dilution effect is to adopt a sample
of source galaxies that are located behind all the clusters of
interest. As discussed in \citet{medezinski17}, we can construct a
secure background galaxy sample by applying cuts in the color-color
space or by taking advantage of the probability distribution functions
of photometric redshifts for individual source galaxies. Most of shear
selected clusters are located at $z\lesssim 0.7$, which implies that
mass map peaks from a source galaxy sample that contains only galaxies
at $z\gtrsim 0.7$ are immune to the dilution effect, although such
selection reduces the number density of source galaxies and therefore
reduces the number of mass map peaks. This can be seen as a trade-off
between the statistical and systematic errors. We explore the effect of
cluster member galaxies in more details in future work.

\end{document}

%% file: papertablephotozmodclimit.tex
1 &  8.62 & 245.3800 &  42.7656 & 0.13 &  0.152 &  33.0 &  2.02$\pm$ 0.41 &  7.4$\pm$ 4.6 & Abell 2183 (z=0.1365) (W)\\ 
2 &  7.52 &  37.9254 &  -4.8803 & 0.05 &  0.186 & 116.4 &  2.72$\pm$ 0.48 &  2.9$\pm$ 1.1 & Abell 0362 (z=0.1843) (W)\\ 
3 &  7.47 & 336.0460 &   0.3360 & 0.11 &  0.154 &  48.9 &  1.83$\pm$ 0.39 &  7.5$\pm$ 5.2 &  (W)\\ 
4 &  7.25 & 138.4610 &  -0.7610 & 0.25 &  0.284 &  36.1 &  2.96$\pm$ 0.69 &  2.1$\pm$ 1.0 &  (W)\\ 
5 &  7.02 & 179.0520 &  -0.3463 & 0.33 &  0.254 &  66.2 &  3.43$\pm$ 0.67 &  1.5$\pm$ 0.6 &  (W)\\ 
6 &  6.41 & 338.9120 &   1.4802 & -- &  0.057$^\dagger$ & -- &  1.73$\pm$ 0.43  &  1.6$\pm$ 0.8  & Abell 2457 (z=0.0594) (W)\\ 
7 &  6.40 & 133.6610 &   0.6372 & 0.50 &  0.121 &  28.3 &  1.21$\pm$ 0.31 & 10.0$\pm$ 0.0 &  (W)\\ 
8 &  6.39 & 139.0430 &  -0.3948 & 0.33 &  0.318 &  92.3 &  2.96$\pm$ 0.81 &  4.3$\pm$ 2.5 & Abell 0776 (z=0.3359) (W)\\ 
9 &  6.36 &  37.3951 &  -3.6099 & 0.11 &  0.312 &  57.0 &  4.16$\pm$ 0.86 &  1.3$\pm$ 0.5 &  (W)\\ 
10 &  6.30 & 177.5860 &  -0.6043 & 0.08 &  0.135 &  51.9 &  1.51$\pm$ 0.33 &  4.5$\pm$ 2.6 &  (W)\\ 
11 &  6.20 & 221.0390 &   0.1773 & -- &  0.294$^\dagger$ & -- &  3.38$\pm$ 0.75  &  1.4$\pm$ 0.6  & (W)\\ 
12 &  6.17 & 180.4270 &  -0.1880 & 0.16 &  0.164 &  48.0 &  1.69$\pm$ 0.45 &  1.5$\pm$ 0.8 & Abell 1445 (z=0.1694)\\ 
13 &  5.99 &  30.4210 &  -5.0203 & 0.93 &  0.809 &  24.0 & -- & -- & \\ 
  &  &  &  & 0.12 &  0.206 &  15.4 & -- & -- & \\ 
14 &  5.99 & 216.7760 &   0.7232 & 0.04 &  0.296 &  25.8 &  2.66$\pm$ 0.63 &  1.5$\pm$ 0.7 &  (W)\\ 
15 &  5.95 & 139.7060 &   2.2068 & 0.41 &  0.298 &  24.2 &  2.71$\pm$ 0.61 &  5.6$\pm$ 3.5 &  (W)\\ 
16 &  5.77 &  33.3622 &  -2.9150 & 0.40 &  0.150 &  34.7 &  1.29$\pm$ 0.42 &  2.1$\pm$ 1.5 &  (W)\\ 
17 &  5.74 & 178.0610 &   0.5243 & 0.23 &  0.472 &  62.3 &  3.50$\pm$ 0.99 &  7.2$\pm$ 5.2 &  (W)\\ 
18 &  5.71 & 133.1200 &   0.4081 & 0.40 &  0.270 &  44.3 &  1.92$\pm$ 0.48 & 10.0$\pm$ 0.0 &  (W)\\ 
19 &  5.64 & 177.1120 &  -0.6583 & 1.21 &  0.420 &  28.0 &  2.41$\pm$ 0.89 &  1.5$\pm$ 1.2 & \\ 
20 &  5.55 & 138.5050 &   1.6646 & 1.48 &  0.380 &  16.9 &  3.67$\pm$ 1.09 &  1.9$\pm$ 1.3 &  (W)\\ 
21 &  5.45 & 336.2300 &  -0.3638 & 0.18 &  0.140 &  19.2 & -- & -- & \\ 
  &  &  &  & 0.17 &  0.307 &  33.8 & -- & -- & \\ 
22 &  5.43 & 130.3710 &   0.4399 & 1.43 &  0.454 &  25.8 & -- & -- & \\ 
  &  &  &  & 0.85 &  0.215 &  24.9 & -- & -- & \\ 
  &  &  &  & 0.12 &  0.413 &  22.0 & -- & -- & \\ 
23 &  5.42 & 213.8900 &  -0.0515 & -- & -- & -- & -- & --  & \\ 
24 &  5.39 & 245.0570 &  42.5066 & 0.16 &  0.142 &  27.9 &  2.17$\pm$ 0.83 &  0.4$\pm$ 0.2 &  (W)\\ 
25 &  5.37 & 219.2130 &  -0.7013 & 0.80 &  0.198 &  18.3 &  0.88$\pm$ 0.28 & 10.0$\pm$ 0.0 &  (W)\\ 
26 &  5.37 & 213.7780 &  -0.4879 & 0.08 &  0.144 &  43.0 &  1.97$\pm$ 0.67 &  0.4$\pm$ 0.2 & Abell 1882 (z=0.1367) (W)\\ 
27 &  5.35 & 134.1180 &   2.1991 & -- &  0.126$^\dagger$ & -- &  0.00$\pm$ 0.00  &  0.0$\pm$ 0.0  & (W)\\ 
28 &  5.33 &  38.1165 &  -4.7920 & 0.96 &  0.276 &  33.7 &  1.68$\pm$ 0.51 &  2.5$\pm$ 1.8 &  (W)\\ 
29 &  5.33 &  30.3810 &  -5.5104 & -- &  0.194$^\dagger$ & -- &  1.36$\pm$ 0.37  & 10.0$\pm$ 0.0  & (W)\\ 
30 &  5.32 & 140.6830 &   2.1374 & 0.07 &  0.194 &  24.7 &  1.15$\pm$ 0.33 & 10.0$\pm$ 0.0 &  (W)\\ 
31 &  5.31 & 219.4300 &  -0.3264 & 0.68 &  0.147 &  51.3 &  1.98$\pm$ 0.56 &  1.1$\pm$ 0.6 &  (W)\\ 
32 &  5.27 & 178.1020 &  -0.5087 & 0.64 &  0.311 &  15.7 &  2.46$\pm$ 0.77 &  1.2$\pm$ 0.7 &  (W)\\ 
33 &  5.22 & 215.0660 &   0.9680 & 0.84 &  0.168 &  16.6 & -- & -- & \\ 
  &  &  &  & 0.45 &  0.515 &  46.3 & -- & -- & \\ 
  &  &  &  & 0.19 &  0.322 &  16.2 & -- & -- & \\ 
34 &  5.22 & 223.0820 &   0.1644 & 0.18 &  0.592 &  53.8 &  9.59$\pm$ 2.29 &  1.7$\pm$ 0.7 &  (W)\\ 
35 &  5.17 & 211.9870 &  -0.4783 & 0.48 &  0.469 &  35.9 &  3.62$\pm$ 1.18 &  1.4$\pm$ 0.9 &  (W)\\ 
36 &  5.14 &  36.3804 &  -4.2540 & 0.15 &  0.155 &  18.2 &  1.21$\pm$ 0.34 &  3.7$\pm$ 2.7 & \\ 
37 &  5.13 & 240.3980 &  42.7484 & -- &  0.223$^\dagger$ & -- &  1.98$\pm$ 0.47  &  8.0$\pm$ 5.7  & (W)\\ 
38 &  5.08 & 182.4740 &  -0.5622 & 0.39 &  0.178 &  22.4 &  1.64$\pm$ 0.56 &  4.8$\pm$ 4.6 &  (W)\\ 
39 &  5.07 & 140.4190 &  -0.2530 & 0.56 &  0.310 &  29.7 &  3.08$\pm$ 1.10 &  0.7$\pm$ 0.4 &  (W)\\ 
40 &  5.06 & 216.8680 &  -0.2016 & -- & -- & -- & -- & --  & \\ 
41 &  5.05 & 212.9210 &   0.4053 & 0.33 &  0.252 &  27.1 & -- & -- & \\ 
  &  &  &  & 0.15 &  0.148 &  17.5 & -- & -- & \\ 
42 &  5.04 & 223.0940 &  -0.9723 & 0.07 &  0.304 &  39.0 &  2.56$\pm$ 0.65 &  6.9$\pm$ 5.5 &  (W)\\ 
43 &  5.03 &  37.6634 &  -4.9982 & 0.11 &  0.272 &  33.1 &  1.72$\pm$ 0.66 &  1.4$\pm$ 1.0 &  (W)\\ 
44 &  5.02 & 333.0540 &  -0.1345 & 0.14 &  0.350 &  30.5 &  3.04$\pm$ 0.88 &  1.5$\pm$ 0.9 &  (W)\\ 
45 &  5.00 & 140.5350 &  -0.4895 & 0.52 &  0.305 &  59.4 &  4.10$\pm$ 1.32 &  1.0$\pm$ 0.6 &  (W)\\ 
46 &  5.00 & 178.7400 &  -1.4377 & 0.14 &  0.158 &  26.5 &  2.67$\pm$ 0.86 &  3.0$\pm$ 3.2 &  (W)\\ 
47 &  5.00 & 217.6800 &   0.8108 & 0.78 &  0.149 &  15.8 & -- & -- & \\ 
  &  &  &  & 0.14 &  0.312 &  36.9 & -- & -- & \\ 
48 &  4.98 & 220.7900 &   1.0509 & 0.33 &  0.529 &  45.8 &  7.30$\pm$ 2.06 &  1.5$\pm$ 0.8 &  (W)\\ 
49 &  4.92 & 223.4620 &  -1.1209 & -- &  0.277$^\dagger$ & -- &  1.72$\pm$ 0.46  & 10.0$\pm$ 0.0  & (W)\\ 
50 &  4.92 & 333.3700 &  -0.1554 & 1.18 &  0.357 &  34.1 & -- & -- & \\ 
  &  &  &  & 0.15 &  0.100 &  22.6 & -- & -- & \\ 
51 &  4.91 & 180.4100 &  -0.5046 & 0.77 &  0.322 &  25.3 & -- & -- & \\ 
  &  &  &  & 0.37 &  0.162 &  23.9 & -- & -- & \\ 
52 &  4.90 & 336.4170 &   1.0777 & 0.31 &  0.281 &  54.4 &  1.74$\pm$ 0.59 &  2.2$\pm$ 1.6 &  (W)\\ 
53 &  4.88 & 220.3950 &  -0.9047 & 1.34 &  0.884 &  16.7 & -- & -- & \\ 
  &  &  &  & 0.24 &  0.534 &  56.7 & -- & -- & \\ 
54 &  4.87 & 213.6040 &  -0.3631 & 0.33 &  0.144 &  68.8 &  1.69$\pm$ 0.68 &  0.2$\pm$ 0.1 &  (W)\\ 
55 &  4.84 & 224.2570 &  -1.0034 & -- &  0.393$^\dagger$ & -- &  2.78$\pm$ 0.91  &  9.4$\pm$11.1  & (W)\\ 
56 &  4.83 & 335.4000 &   1.3859 & 1.08 &  0.790 &  23.4 & -- & -- & \\ 
  &  &  &  & 0.94 &  0.324 &  17.6 & -- & -- & \\ 
57 &  4.82 & 221.3300 &   0.1107 & 0.09 &  0.287 &  34.4 &  2.27$\pm$ 0.66 &  1.3$\pm$ 0.7 &  (W)\\ 
58 &  4.78 &  33.1330 &  -5.5513 & 0.97 &  0.287 &  64.6 &  1.84$\pm$ 0.81 &  0.7$\pm$ 0.5 & \\ 
59 &  4.77 & 338.0160 &   0.0277 & -- & -- & -- & -- & --  & \\ 
60 &  4.75 & 180.6600 &  -1.3542 & 0.14 &  0.246 &  29.7 &  1.44$\pm$ 0.48 &  5.2$\pm$ 5.0 &  (W)\\ 
61 &  4.73 & 337.1260 &   1.7066 & 0.09 &  0.338 &  36.0 &  2.58$\pm$ 0.62 & 10.0$\pm$ 0.0 &  (W)\\ 
62 &  4.73 & 216.6510 &   0.7982 & -- &  0.592$^\dagger$ & -- &  7.17$\pm$ 2.71  &  0.4$\pm$ 0.2  & (W)\\ 
63 &  4.71 &  33.4748 &  -2.8818 & 1.17 &  0.288 &  22.3 &  1.28$\pm$ 0.53 &  2.1$\pm$ 2.0 & \\ 
64 &  4.71 & 220.5860 &   0.3355 & 0.23 &  0.172 &  19.2 &  0.95$\pm$ 0.29 & 10.0$\pm$ 0.0 &  (W)\\ 
65 &  4.70 & 339.7630 &   0.6681 & 0.74 &  0.200 &  19.2 & -- & -- & \\ 
  &  &  &  & 0.26 &  0.264 &  21.3 & -- & -- & \\ 

%% file: ms.bbl
\begin{thebibliography}{}
\bibitem[Aihara et al.(2017)]{aihara17}
Aihara, H., et al.\ 2017, arXiv:1704.05858
\bibitem[Anderson et al.(2015)]{Anderson:2015} 
Anderson M.~E., Gaspari M., White S.~D.~M., Wang W., Dai X., 2015, MNRAS, 449, 3806
\bibitem[Axelrod et al.(2010)]{lsst-stack}
Axelrod, T., Kantor, J., Lupton, R.~H., \& Pierfederici, F.\ 2010,
\procspie, 7740, 774015  
\bibitem[Berlind et al.(2006)]{berlind06}
Berlind, A.~A., et al.\ 2006, \apjs, 167, 1
\bibitem[Balogh \& McGee(2010)]{balogh10}
Balogh, M.~L., \& McGee, S.~L.\ 2010, \mnras, 402, L59 
\bibitem[Becker et al.(2011)]{becker11}
Becker, M.~R., \& Kravtsov, A.~V.\ 2011, \apj, 740, 25
\bibitem[Bhattacharya et al.(2013)]{bhattacharya13}
Bhattacharya, S., Habib, S., Heitmann, K., \& Vikhlinin, A.\ 2013, \apj, 766, 32
\bibitem[B{\"o}hringer et al.(2014)]{2014A&A...570A..31B}
B{\"o}hringer, H., Chon, G., \& Collins, C.~A.\ 2014, \aap, 570, A31 
\bibitem[Boller et al.(2016)]{boller16}
Boller, T., Freyberg, M.~J., Tr{\"u}mper, J., Haberl, F., 
Voges, W., \& Nandra, K.\ 2016, \aap, 588, A103 
\bibitem[Bosch et al.(2017)]{bosch17}
Bosch, J., et al.\ 2017, arXiv:1705.06766
\bibitem[Bruzual \& Charlot(2003)]{bruzual03}
Bruzual, G., \& Charlot, S.\ 2003, \mnras, 344, 1000 
\bibitem[Diemer \& Kravtsov(2015)]{diemer15}
Diemer, B., \& Kravtsov, A.~V.\ 2015, \apj, 799, 108 
\bibitem[Dietrich et al.(2012)]{dietrich12}
Dietrich, J.~P. et al.\ 2012 \mnras, 419, 3547
\bibitem[Duffy et al.(2008)]{duffy08}
Duffy, A.~R., Schaye, J., Kay, S.~T., \& Dalla Vecchia, C.\ 2008, 
\mnras, 390, L64 
\bibitem[Dutton \& Macci{\`o}(2014)]{dutton14}
Dutton, A.~A., \& Macci{\`o}, A.~V.\ 2014, \mnras, 441, 3359 
\bibitem[Filho et al.(2014)]{filho14}
Filho, M.~E., Brochado, P., Brinchmann, J., Lobo, C., Henriques, B.,
Gr\"{u}tzbauch, R., \& Gomes, J.~M.\ 2014, \mnras, 443, 288
\bibitem[Ford \& VanderPlas(2016)]{ford16}
Ford, J., \& VanderPlas, J.\ 2016, \aj, 152, 228 
\bibitem[Gavazzi \& Soucail(2007)]{2007A&A...462..459G}
Gavazzi, R., \& Soucail, G.\ 2007, \aap, 462, 459 
\bibitem[Gehrels (1986)]{gehrels86}
Gehrels, N. 1986, \apj, 303, 336
\bibitem[Giles et al.(2015)]{2015MNRAS.447.3044G}
Giles, P.~A., Maughan, B.~J., Hamana, T., Miyazaki, S., 
Birkinshaw, M., Ellis, R.~S., \& Massey, R.\ 2015, \mnras, 447, 3044 
\bibitem[Gonzalez et al.(2013)]{gonzalez13}
Gonzalez, A.~H., Sivanandam, S., Zabludoff, A.~I., 
\& Zaritsky, D.\ 2013, \apj, 778, 14 
\bibitem[Goto et al.(2002)]{goto02}
Goto, T., et al.\ 2002, \aj, 123, 1807
\bibitem[Hamana et al.(2004)]{hamana04}
Hamana, T., Takada, M., \& Yoshida, N.\ 2004, \mnras, 350, 893 
\bibitem[Hamana et al.(2012)]{2012MNRAS.425.2287H}
Hamana, T., Oguri, M., Shirasaki, M., \& Sato, M.\ 2012, \mnras, 425, 2287 
\bibitem[Hamana et al.(2015)]{2015PASJ...67...34H}
Hamana, T., Sakurai, J., Koike, M., \& Miller, L.\ 2015, \pasj, 67, 34 
\bibitem[Hennawi \& Spergel(2005)]{henna05}
Hennawi, J.~F., \& Spergel, D.~N.\ 2005, \apj, 624, 59 
\bibitem[Hinshaw et al.(2013)]{2013ApJS..208...19H}
 Hinshaw, G., et al.\ 2013, \apjs, 208, 19 
\bibitem[Hirata \& Seljak(2003)]{hirata03}
Hirata, C., \& Seljak, U.\ 2003, \mnras, 343, 459 
\bibitem[Huang et al.(2017)]{huang17}
Huang, S. et al. \ 2017, in this volume
\bibitem[Ivezi\'{c} et al. (2008)]{lsst-stack2}
Ivezi\'{c}, \'{Z}., et al.\ 2008, arXiv:0805.2366\\
(http://www.lsst.org/files/docs/LSSToverview.pdf)
\bibitem[Jain \& van Waerbeke(2000)]{jain00}
Jain, B., \& van Waerbeke, L.\ 2000, \apj, 530, L1
\bibitem[Kaiser \& Squires(1993)]{1993ApJ...404..441K}
Kaiser, N., \& Squires, G.\ 1993, \apj, 404, 441 
\bibitem[Kacprzak et al.(2016)]{2016MNRAS.463.3653K}
Kacprzak, T., et al.\ 2016, \mnras, 463, 3653 
\bibitem[Kushino et al.(2002)]{kushino02}
 Kushino, A., Ishisaki, Y., Morita, U., Yamasaki, N.~Y., Ishida, M., 
Ohashi, T., \& Ueda, Y.\ 2002, \pasj, 54, 327 
\bibitem[Lieu et al.(2016)]{xxl4}
Lieu, M., et al.\ 2016, \aap, 592, A4 
\bibitem[Lin et al.(2004)]{lin04}
Lin, Y.-T., Mohr, J.~J., \& Stanford, S.~A.\ 2004, \apj, 610, 745 
\bibitem[Lin et al.(2006)]{lin06}
Lin, Y.-T., Mohr, J.~J., Gonzalez, A.~H., \& Stanford, S.~A.\ 2006, \apjl, 650,
L99 
\bibitem[Lin et al.(2012)]{lin12}
Lin, Y.-T., Stanford, S.~A., Eisenhardt, P.~R.~M. et al.\ 2012, \apjl, 745, L3
\bibitem[Lin et al.(2017)]{lin17}
Lin, Y.-T. et al. 2017, in this volume
\bibitem[Lin et al.(2016)]{2016A&A...593A..88L}
Lin, C.-A., Kilbinger, M., \& Pires, S.\ 2016, \aap, 593, A88 
\bibitem[Liu et al.(2015)]{2015MNRAS.450.2888L}
Liu, X., et al.\ 2015, \mnras, 450, 2888 
\bibitem[Liu et al.(2015)]{2015PhRvD..91f3507L}
Liu, J., Petri, A., Haiman, Z., Hui, L., Kratochvil, J.~M., \& May,
M.\ 2015, \prd, 91, 063507  
\bibitem[Ludlow et al.(2010)]{ludlow10}
Ludlow, A.~D., Navarro, J.~F., Springel, V., Vogelsberger, M., 
Wang, J., White, S.~D.~M., Jenkins, A., \& Frenk, C.~S.\ 2010, \mnras, 406, 137 
\bibitem[Mandelbaum et al.(2006)]{2006MNRAS.372..758M}
Mandelbaum, R., Seljak, U., Cool, R.~J., Blanton, M., 
Hirata, C.~M., \& Brinkmann, J.\ 2006, \mnras, 372, 758 
\bibitem[Mandelbaum et al.(2017)]{mandelbaum17}
Mandelbaum, R., et al.\ 2017, arXiv:1705.06745
\bibitem[Marian \& Bernstein(2006)]{2006PhRvD..73l3525M}
Marian, L., \& Bernstein, G.~M.\ 2006, \prd, 73, 123525 
\bibitem[Marian et al.(2010)]{2010ApJ...709..286M}
Marian, L., Smith, R.~E., \& Bernstein, G.~M.\ 2010, \apj, 709, 286 
\bibitem[Marian et al.(2012)]{2012MNRAS.423.1711M}
Marian, L., Smith, R.~E., Hilbert, S., \& Schneider, P.\ 2012, \mnras, 423, 1711 
\bibitem[Maturi et al.(2005)]{2005A&A...442..851M}
Maturi, M., Meneghetti, M., Bartelmann, M., Dolag, K., \& Moscardini,
L.\ 2005, \aap, 442, 851  
\bibitem[Medezinski et al.(2010)]{medezinski10}
Medezinski, E., et al.\ 2010, \mnras, 405, 257
\bibitem[Medezinski et al.(2017)]{medezinski17}
Medezinski, E., et al.\ 2017, arXiv:1706.00427
\bibitem[Miyazaki et al.(2007)]{2007ApJ...669..714M}
Miyazaki, S., Hamana, T., Ellis, R.~S., Kashikawa, N., Massey, R.~J.,
Taylor, J., \& Refregier, A.\ 2007, \apj, 669, 714 

\bibitem[Miyazaki et al.(2015)]{2015ApJ...807...22M}
Miyazaki, S., et al.\ 2015, \apj, 807, 22
\bibitem[Miyazaki et al.(2017)]{miyazaki17}
Miyazaki, S., et al.\ 2017, \pasj, in press
\bibitem[Navarro et al.(1997)]{navarro97}
Navarro, J.~F., Frenk, C.~S., \& White, S.~D.~M.\ 1997, \apj, 490, 493 
\bibitem[Oguri(2014)]{oguri14}
Oguri, M.\ 2014, \mnras, 444, 147 
\bibitem[Oguri et al.(2012)]{2012MNRAS.420.3213O}
Oguri, M., Bayliss, M.~B., Dahle, H., Sharon, K., Gladders, M.~D.,
Natarajan, P., Hennawi, J.~F., \& Koester, B.~P., 2012, \mnras, 420, 3213 
\bibitem[Oguri \& Hamana(2011)]{ogurihamana11}
Oguri, M., Hamana, T. 2011, \mnras, 414, 1851
\bibitem[Oguri et al.(2017a)]{oguri17a}
Oguri, M., et al.\ 2017a, \pasj, in press (arXiv:1701.00818)
\bibitem[Oguri et al.(2017b)]{oguri17b}
Oguri, M., et al.\ 2017b, arXiv:1705.06792
\bibitem[Okabe \& Smith(2016)]{okabe16}
Okabe, N., Smith, G.~P.\
2016, \mnras, 461, 3794
\bibitem[Pacaud et al.(2016)]{xxl2}
Pacaud, F., et al.\ 2016, \aap, 592, A2 
\bibitem[Pierre et al.(2016)]{xxl1}
Pierre, M., et al.\ 2016, \aap, 592, A1 
\bibitem[Piffaretti et al.(2011)]{2011A&A...534A.109P}
Piffaretti, R., Arnaud, M., Pratt, G.~W., Pointecouteau, E., 
		\& Melin, J.-B.\ 2011, \aap, 534, A109
\bibitem[Planck Collaboration et al.(2016a)]{2016A&A...594A..13P}
Planck Collaboration,  et al.\ 2016, \aap, 594, A13 
\bibitem[Planck Collaboration et al.(2016b)]{2016A&A...594A..24P}
Planck Collaboration,  et al.\ 2016, \aap, 594, A24 
\bibitem[Prada et al.(2012)]{prada12}
Prada, F., Klypin, A.~A., Cuesta, A.~J., Betancort-Rijo, J.~E., 
\& Primack, J.\ 2012, \mnras, 423, 3018 
\bibitem[Schirmer et al.(2007)]{2007A&A...462..875S}
Schirmer, M., Erben, T., Hetterscheidt, M., \& Schneider, P.\ 2007, 
\aap, 462, 875
\bibitem[Schneider(1996)]{1996MNRAS.283..837S}
Schneider, P.\ 1996, \mnras, 283, 837 
\bibitem[Seitz \& Schneider(1995)]{seitz95}
Seitz, C., \& Schneider, P.\ 1995, \aap, 297, 287 
\bibitem[Shan et al.(2012)]{2012ApJ...748...56S}
Shan, H., et al.\ 2012, \apj, 748, 56 
\bibitem[Smith et al.(2001)]{smith01}
Smith, R.~K., Brickhouse, N.~S., Liedahl, D.~A., \& Raymond,
J.~C.\ 2001, \apjl, 556, L91  
\bibitem[Smith et al.(2003)]{smith03}
Smith, R.~E., et al.\ 2003, \mnras, 341, 1311 
\bibitem[Takahashi et al.(2012)]{takahashi12}
Takahashi, R., Sato, M., Nishimichi, T., Taruya, A., \& 
Oguri, M.\ 2012, \apj, 761, 152 
\bibitem[Takahashi et al.(2017)]{takahashi17}
Takahashi, R., Hamana, T., Shirasaki, M., Namikawa, T., Nishimichi,
T., Osato, K., \& Shiroyama, K.\ 2017, arXiv:1706.01472 
\bibitem[Tanaka et al.(2017)]{tanaka17}
Tanaka, M., et al.\ 2017, arXiv:1704.05988
\bibitem[Thanjavur et al.(2009)]{2009ApJ...706..571T}
Thanjavur, K., Willis, J., \& Crampton, D.\ 2009, \apj, 706, 571 
\bibitem[Tinker et al.(2008)]{tinker08}
Tinker, J., et al.\ 2008, \apj, 688, 709
\bibitem[Umetsu \& Diemer(2017)]{umetsu17}
Umetsu, K., \& Diemer, B.\ 2017, \apj, 836, 231 
\bibitem[Umetsu et al.(2016)]{umetsu16}
Umetsu, K., et al.\ 20176, \apj, 821, 116
\bibitem[Umetsu et al.(2009)]{umetsu09}
Umetsu, K., et al.\ 2009, \apj, 694, 1643 
\bibitem[Umetsu et al.(2011)]{umetsu11}
Umetsu, K., Broadhurst, T., Zitrin, A., Medezinski, E., 
Coe, D., \& Postman, M.\ 2011, \apj, 738, 41
\bibitem[Utsumi et al.(2014)]{utsumi14}
Utsumi, Y., Miyazaki, S. Geller, M.~J., Dell'Antonio, I.~P., Oguri,
M., Kurtz, M.~J., Hamana, T., \& Fabricant, D.~G.\ 2014, \apj, 786, 93
\bibitem[Voges et al.(1999)]{1999A&A...349..389V}
Voges, W., et al.\ 1999, \aap, 349, 389 
\bibitem[Weinberg \& Kamionkowski(2002)]{weinberg02}
Weinberg, N.~N., \& Kamionkowski, M.\ 2002, \mnras, 337, 1269 
\bibitem[Wen \& Han(2015)]{whl15}
Wen, Z.~L., \& Han, J.~L.\ 2015, \apj, 807, 178 
\bibitem[White et al.(2002)]{white02}
White, M., van Waerbeke, L., \& Mackey, J.\ 2002, \apj, 575, 640 
\bibitem[Wittman et al.(2001)]{2001ApJ...557L..89W}
Wittman, D., Tyson, J.~A., Margoniner, V.~E., Cohen, J.~G., \& 
Dell'Antonio, I.~P.\ 2001, \apjl, 557, L89 
\bibitem[Wittman et al.(2006)]{2006ApJ...643..128W} 
Wittman, D., Dell'Antonio, I.~P., Hughes, J.~P., Margoniner, V.~E.,
Tyson, J.~A., Cohen, J.~G., \& Norman, D.\ 2006, \apj, 643, 128 
\end{thebibliography}
